\documentclass[12pt]{iopart}
\usepackage{iopams}

\usepackage{graphicx}

\begin{document}

\title{Cosmic microwave anisotropies in an inhomogeneous compact flat universe}

\author{R.~Aurich and S.~Lustig}

\address{Institut f\"ur Theoretische Physik, Universit\"at Ulm,\\
Albert-Einstein-Allee 11, D-89069 Ulm, Germany
}

\begin{abstract}
The anisotropies of the cosmic microwave background (CMB) are computed
for the half-turn space $E_2$
which represents a compact flat model of the Universe,
i.\,e.\ one with finite volume.
This model is inhomogeneous in the sense that the statistical properties
of the CMB depend on the position of the observer within the fundamental cell.
It is shown that the half-turn space describes the observed CMB anisotropies
on large scales better than the concordance model with infinite volume.
For most observer positions it matches the temperature correlation function
even slightly better than the well studied 3-torus topology.
\end{abstract}

\pacs{98.80.-k, 98.70.Vc, 98.80.Es}


\section{Introduction}

One of the enigmas of the cosmic microwave background (CMB)
is the low power in the temperature correlations at large angles $\vartheta$.
This behaviour is most clearly revealed by the
temperature 2-point correlation function $C(\vartheta)$
which is defined as
\begin{equation}
\label{Eq:C_theta}
C(\vartheta) \; := \; \left< \delta T(\hat n) \delta T(\hat n')\right>
\hspace{10pt} \hbox{with} \hspace{10pt}
\hat n \cdot \hat n' = \cos\vartheta
\hspace{10pt} ,
\end{equation}
where $\delta T(\hat n)$ is the temperature fluctuation in
the direction of the unit vector $\hat n$.
Already the COBE team \cite{Hinshaw_et_al_1996} discovered
the surprisingly low power at large angles $\vartheta \gtrsim 60^\circ$
which is at variance with the $\Lambda$CDM concordance model
as has been found by \cite{Spergel_et_al_2003}
and recently emphasised by \cite{Aurich_Janzer_Lustig_Steiner_2007,%
Copi_Huterer_Schwarz_Starkman_2008,%
Copi_Huterer_Schwarz_Starkman_2010}.
The reality of this discordance is questioned in
\cite{Efstathiou_Ma_Hanson_2009}
such that it could arise as an artefact of method of analysis. 
The arguments are further investigated in \cite{Aurich_Lustig_2010}
with the conclusion that it is very likely
that the low power at large angles is real.
In the following we take the latter point of view.
Then there arises the desire for an explanation of this suppression of power.

One explanation could be that the universe possesses a non-trivial topology,
i.\,e.\ that the spatial space is multi-connected.
For an introduction in cosmic topology, see
\cite{Lachieze-Rey_Luminet_1995,Luminet_Roukema_1999,Levin_2002,%
Reboucas_Gomero_2004,Luminet_2008}.
In that case the multi-connected space would lead to a natural lower cut-off
in the wave numbers describing the perturbations leading to the
temperature anisotropy in the CMB.
This mechanism works provided that the volume of the fundamental cell
is not larger than the volume within the surface of last scattering.
On the other hand too small volumes are also excluded
since they lead to a too strong suppression of the correlations
also at smaller angular scales.
The predicted CMB anisotropy thus depends on
the size of the fundamental cell.

Multi-connected space forms are possible in all three spaces of
constant curvature, i.\,e.\ in hyperbolic, flat, and spherical spaces.
Since the favoured $\Lambda$CDM concordance model describes our Universe
as a flat space,
we restrict ourselves in the following also to the flat case.
In the Euclidean space ${\mathbb E}^3$
there exist 18 topologically different space forms,
but only 10 possess a finite volume,
from these are four non-orientable
\cite{Lachieze-Rey_Luminet_1995,Riazuelo_et_al_2004}.
The remaining 6 flat models are of great promise
in order to explain the low power in the CMB anisotropy at large scales.
Only one from these six multi-connected spaces possesses the special property
of global homogeneity which means that the statistical properties of the CMB
are independent from the position of the observer.
This well studied case is the 3-torus, also called hyper-torus,
where the three pairs of opposing faces are each identified.
Because of the homogeneity it suffices to compute,
e.\,g.\ the temperature correlation function $C(\vartheta)$
defined in Eq.\,(\ref{Eq:C_theta}) for one observer in the 3-torus.
The ensemble average of $C(\vartheta)$ and its cosmic variance
for this observer is identical to that off all other observers.
This facilitates the numerical analysis and the comparison with
the observational data.

This contrasts to the five remaining inhomogeneous flat space forms
that are orientable and possess a finite volume.
These are called $E_2$ to $E_6$ in
\cite{Lachieze-Rey_Luminet_1995,Riazuelo_et_al_2004}.
This paper puts the focus on the space form $E_2$,
also called half-turn space,
and presents a systematic observer dependent analysis of the
statistical properties of the CMB.
In a pioneering work
\cite{Levin_Scannapieco_Silk_1998,Scannapieco_Levin_Silk_1999}
the statistical properties are investigated for
two positions of the observer
which already reveal the suppression of the large-scale power.
The investigation of the half-turn space is extended
in \cite {Riazuelo_et_al_2004}
where the angular power spectrum $\delta T_l^2$ is shown
for six different positions of the observer
for a single sky realisation.
But again, no systematic analysis is carried out
which is the aim of this paper.


\section{The half-turn space and its eigenmodes}

\begin{figure}[htb]
\begin{center}
{
\vspace*{-10pt}\begin{minipage}{18cm}
\hspace*{20pt}\includegraphics[width=8cm]{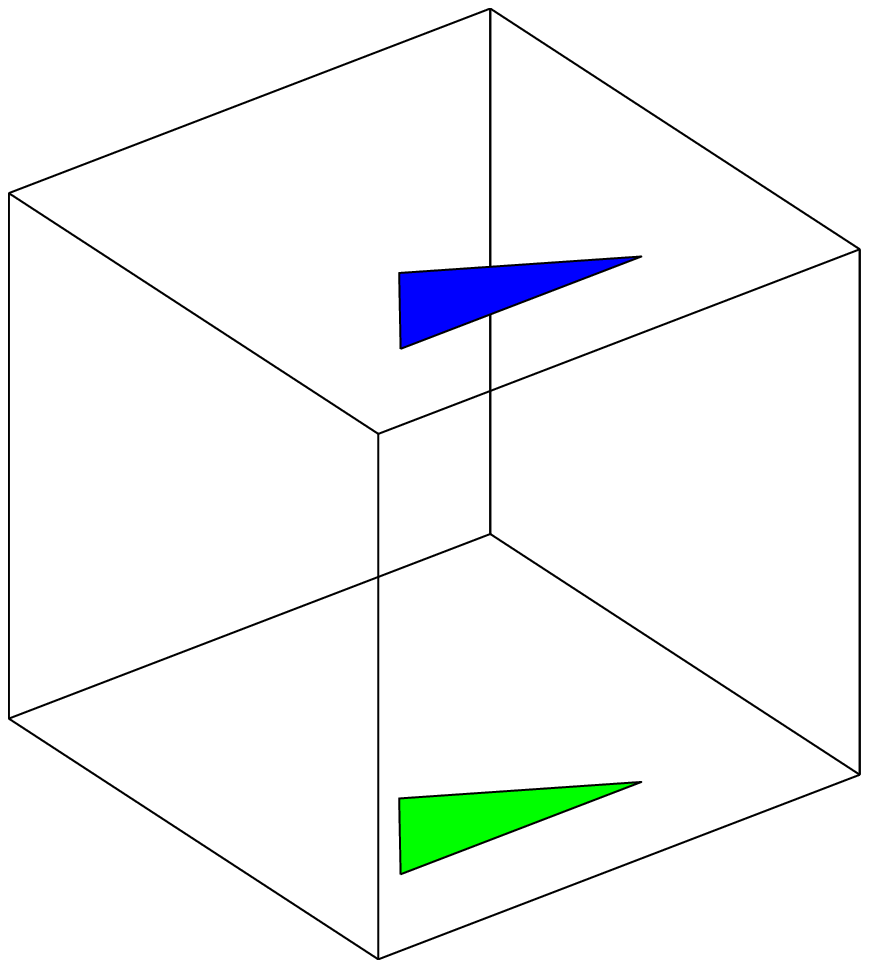}
\hspace*{-40pt}\includegraphics[width=8cm]{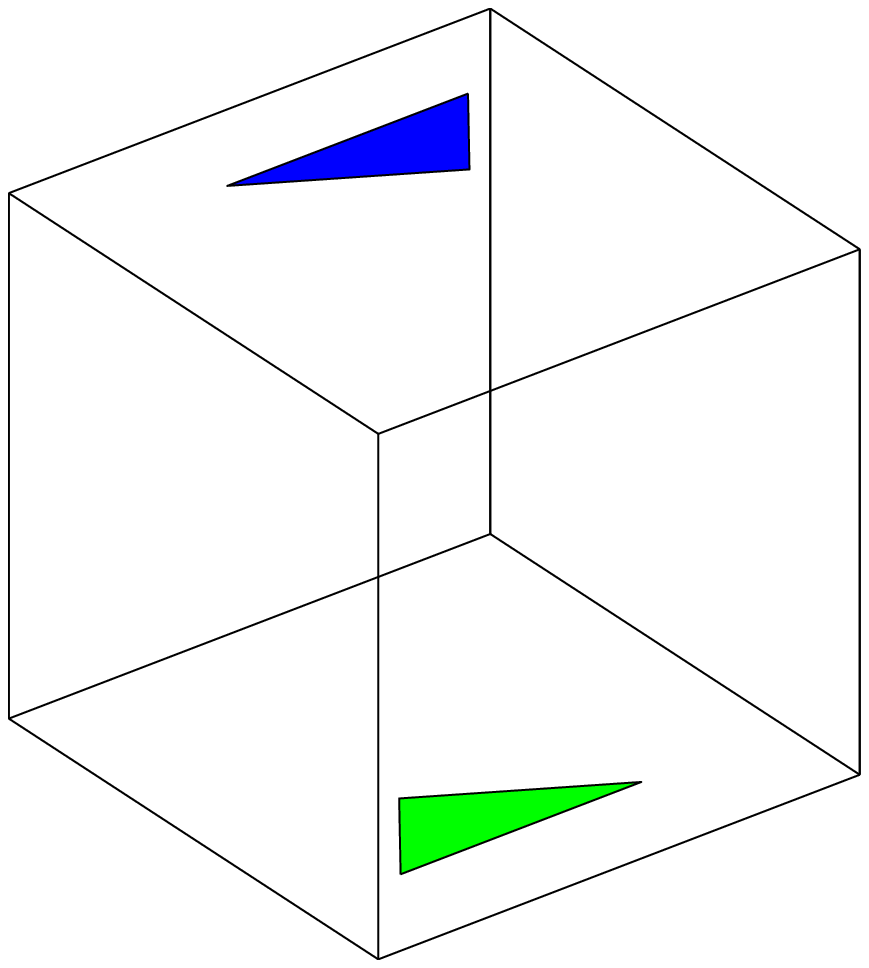}
\end{minipage}
}
\end{center}
\vspace*{-20pt}\caption{
\label{Fig:Fundamental_Cell}
The sketch illustrates the difference between the 3-torus topology (left)
and the half-turn space (right).
The identifications of the vertical faces are the same for both models.
Only the transformation from the bottom face to the top face
involves a rotation in the case of the half-turn space
which is absent for the 3-torus.
}
\vspace*{-5pt}
\end{figure}

The Euclidean space forms are obtained as the quotient
${\mathbb E}^3/\Gamma$ of the Euclidean space ${\mathbb E}^3$
by a discrete and fixed point free symmetry group $\Gamma$.
The simplest model is the 3-torus
in which case the group $\Gamma$ of symmetries is generated by
three orthogonal translations which shift the points
by the lengths $L_x$, $L_y$, and $L_z$.
This model has the special property of homogeneity.
The simplest model without this property,
but which has finite volume and is orientable,
is the half-turn space.
One generator of the 3-torus, say the one in the $z$-direction,
is replaced by a translation accompanied by a rotation by an angle
of $180^\circ$.
The half-turn space is then generated by the three transformations
\begin{eqnarray}
\label{Eq:generator} \nonumber
&\vec{x} \rightarrow \vec{x}\,' \,=\, \vec{x}+L_x\,\vec{e}_x \nonumber\\
&\vec{x} \rightarrow \vec{x}\,' \,=\, \vec{x}+L_y\,\vec{e}_y \\ \nonumber
&\vec{x} \rightarrow \vec{x}\,' \,=\, \vec{x}_{\hbox{\scriptsize R}}+L_z\,\vec{e}_z
\hspace{5pt}
\end{eqnarray}
where $\vec{x}_{\hbox{\scriptsize R}}\,=\,(-x\,L_x,-y\,L_y,z\,L_z)$
takes the rotation of $180^\circ$ in the $xy$-plane into account.
The dimensionless coordinates $x$, $y$, $z \in [-0.5, 0.5]$ allow
the description of points within the fundamental cell
without reference to the topological length scales $L_x$, $L_y$, and $L_z$.
The difference between the 3-torus and the half-turn space is illustrated
in figure \ref{Fig:Fundamental_Cell}
where the $z$-transformation identifies the bottom and the top faces.
In the case of the 3-torus (left) it is a simple shift
whereas for the half-turn space the additional rotation
leads to a further twist as illustrated by the triangle.
This modification leads to an inhomogeneous space form.
For a definition of homogeneous space forms,
see \cite{Ratcliffe_2005} p.\,16
and \cite{Wolf_1974} p.\,135.
All space forms that are not homogeneous are called inhomogeneous.

\begin{figure}[htb]
\begin{center}
{
\vspace*{-10pt}\begin{minipage}{18cm}
\hspace*{52pt}\includegraphics[width=7cm]{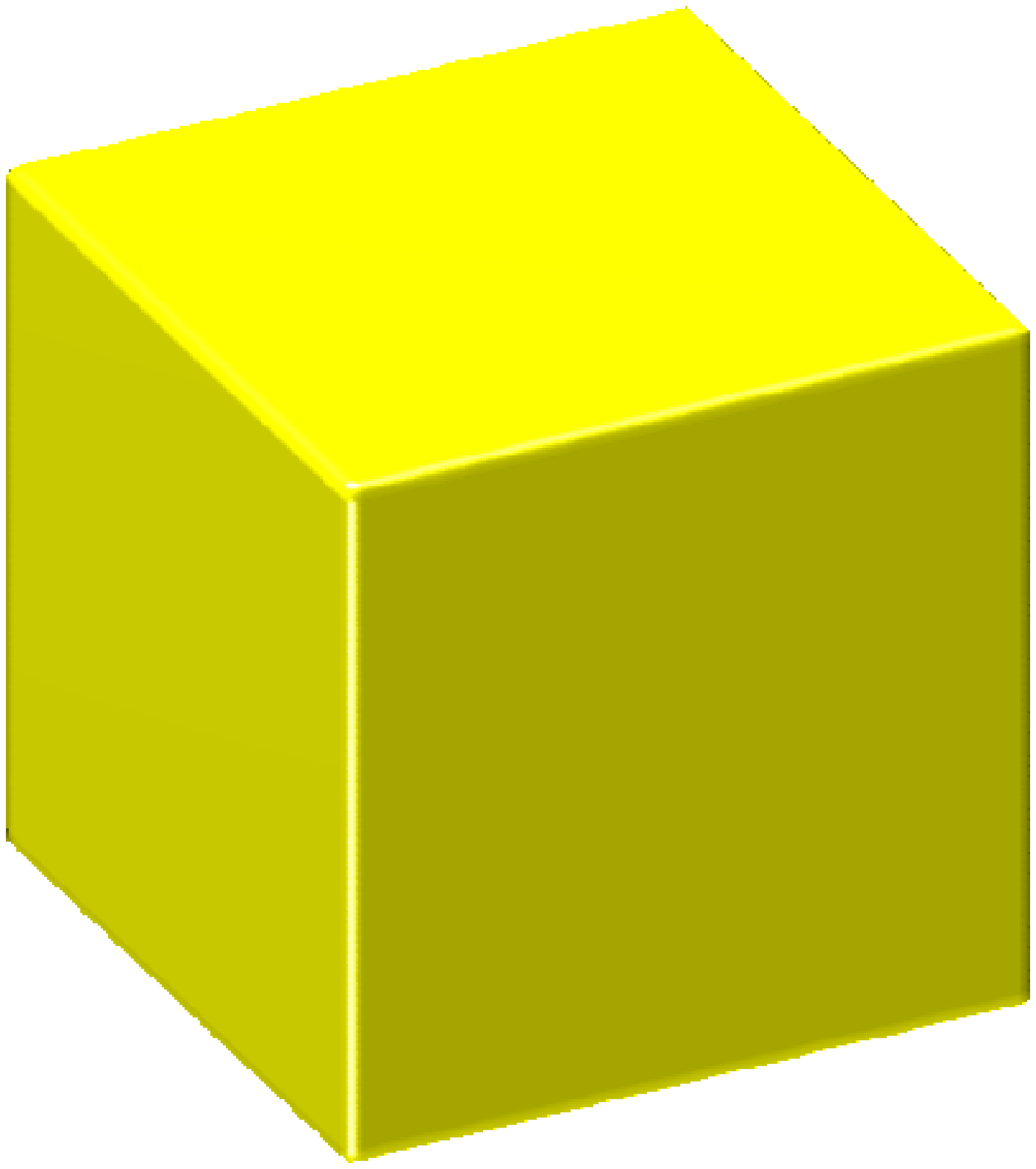}
\hspace*{0pt}\includegraphics[width=7cm]{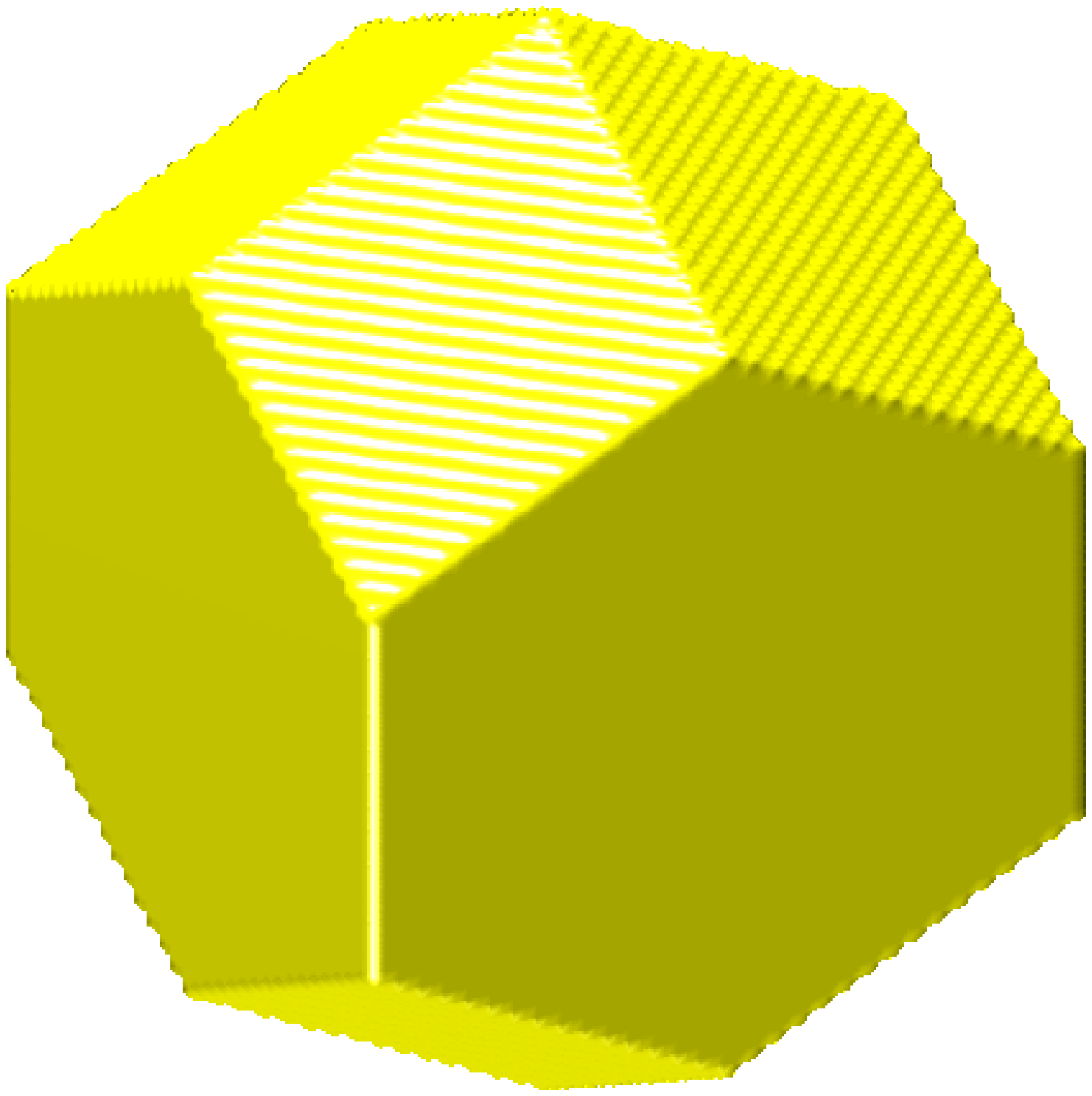}
\end{minipage}
}
\end{center}
\vspace*{-20pt}\caption{
\label{Fig:Inhomogeneous}
The fundamental cell of the half-turn space is shown for the
observer positions $\vec{x}_o=(0,0,0)$ (left)
and $\vec{x}_o=(\frac14\,L_x,\frac14\,L_y,0)$ (right).
Both cells have the same volume.
}
\vspace*{-5pt}
\end{figure}

The inhomogeneity can be visualised by the fundamental cell.
Let us define the fundamental cell with respect to an
observer position $\vec{x}_o$ as the set of points $\vec{x}$
that cannot be transformed closer to $\vec{x}_o$ by applying
any of the transformations $g\in\Gamma$.
For a homogeneous space form like the 3-torus,
the fundamental cell is independent of the observer position $\vec{x}_o$
but not for an inhomogeneous one
as it is illustrated by figure \ref{Fig:Inhomogeneous}.
For the observer position $\vec{x}_o=(0,0,0)$ a cubic fundamental cell
is obtained, whereas for the shifted position
$\vec{x}_o=(\frac14\,L_x,\frac14\,L_y,0)$ a much more complex
fundamental cell is seen by the observer.
Note that the symmetry group $\Gamma$ is the same in both cases.
This behaviour leads to different statistical properties of the CMB.

The eigenfunctions of the Laplacian of the Euclidean space
are plane waves or linear combinations thereof.
In the case of a multi-connected space form every eigenfunction
must be invariant under the action of the generators of the manifold.
This restricts the admissible wave numbers $\vec k$ occurring in
the eigenfunctions
which can be expressed by
$\vec{k}=2\pi(n_x/L_x, n_y/L_y, n_z/2L_z)$ and
$\vec{k}^{\,'}=2\pi(-n_x/L_x, -n_y/L_y, n_z/2L_z)$.
The eigenfunctions of the half-turn space
depend on the values of the integers $n_x\geq 0$, $n_y$, and $n_z$.
For $n_x=n_y=0$, $n_z \in 2\,{\mathbb Z}$ the eigenfunctions are given by
\begin{equation}
\label{Eq:eigenfunction_a}
\Psi_{\vec k}\,(\vec x\,) \, = \,
\exp\left(\hbox{i}\, \vec{k}\cdot \vec{x}\right)
\end{equation}
and for $n_x \in {\mathbb N}$, $n_y$, $n_z \in {\mathbb Z}$
or $n_x=0$, $n_y \in {\mathbb N}$,  $n_z \in {\mathbb Z}$ by 
\begin{equation}
\label{Eq:eigenfunction_b}
\Psi_{\vec k}\,(\vec x\,)\, = \,
\frac{1}{\sqrt{2}}\left[\exp\left(\hbox{i}\, \vec{k}\cdot \vec{x}\right)
+ (-1)^{n_z}\exp\left(\hbox{i}\, \vec{k}^{\,'}\cdot \vec{x}\right) \right]
\hspace{10pt} .
\end{equation}
To normalise the eigenfunctions with respect to the fundamental cell,
they have to be multiplied by $1/\sqrt{L_xL_yL_z}$.
We drop this overall factor in the following,
since the CMB anisotropy has to be normalised with respect to the data.
The computation of the CMB anisotropy requires the expansion
of the eigenfunctions with respect to the spherical basis
\begin{equation}
\label{Eq:eigenfunction_spherical_basis}
\Psi_{\vec{k}}\,(r,\hat{n},\vec{x}_o)\, = \,
\sum_{l,m}\xi_{lm}^{\vec{k}}(\vec{x}_o) R_{k\,l}(r)\,Y_{l\,m}(\hat{n})
\end{equation}
where $R_{k\,l}(r)=4\pi\,j_l(kr)$ is the radial function,
i.\,e.\ the spherical Bessel function,
$Y_{l\,m}(\hat{n})$ the spherical harmonics,  
$r=\left|\vec{x}-\vec{x}_o\right|$,
$\hat{n}=\left(\vec{x}-\vec{x}_o\right)/r$, $\vec{x}_o$
the position of the observer,
and $k=|\vec k\, | = |\vec k ^{\,'}|$.
The expansion coefficients $\xi_{lm}^{\vec{k}}(\vec{x}_o)$
for Eq.\,(\ref{Eq:eigenfunction_a}) are given by
\begin{equation}
\label{Eq:spherical_expansion_a}
\xi_{lm}^{\vec{k}}\,(\vec{x}_o)\,=\,
\hbox{i}^l \, Y_{lm}^{*}(\hat{k})\,\exp\left(\hbox{i} \, \vec{k}\cdot  
\vec{x}_o\right)
\end{equation}
and for Eq.\,(\ref{Eq:eigenfunction_b})
\begin{eqnarray}
\label{Eq:spherical_expansion_b}
\xi_{lm}^{\vec{k}}\,(\vec{x}_o) &=&
\frac{\hbox{i}^l}{\sqrt{2}}\,
\left[Y_{lm}^{*}(\hat{k})\,\exp\left(\hbox{i} \, \vec{k}\cdot  
\vec{x}_o\right)
+ (-1)^{n_z}\,Y_{lm}^{*}(\hat{k}^{\,'})\,\exp\left(\hbox{i} \,
\vec{k}^{\,'}\cdot \vec{x}_o\right)\right]
\nonumber\\ & = &
\frac{\hbox{i}^l}{\sqrt{2}}\,Y_{lm}^{*}(\hat{k})\,
\left[\exp\left(\hbox{i} \, \vec{k}\cdot \vec{x}_o\right)
+ (-1)^{n_z+m}\,\exp\left(\hbox{i} \,
\vec{k}^{\,'}\cdot \vec{x}_o\right)\right]
\end{eqnarray}
where $Y_{lm}^{*}(\hat{k}^{\,'})=(-1)^mY_{lm}^{*}(\hat{k})$,
$\hat{k}=\vec{k}/k$, and $\hat{k}^{\,'}=\vec{k}^{\,'}/k$.

Expanding the temperature fluctuations of the CMB 
according to the spherical harmonics, i.\,e.
\begin{equation}
\label{Eq:spherical_expansion_dT}
\delta T(\hat{n})=\sum_{l,m} a_{lm}\,Y_{lm}(\hat{n})
\hspace{10pt} ,
\end{equation}
the corresponding coefficients $a_{lm}$ of the half-turn space
are determined by
\begin{equation}
\label{Eq:a_lm_half_turn_space}
a_{lm} \; = \;
\sum_{\vec{k}}T_l(k) \; \Phi_{\vec{k}} \; \xi_{lm}^{\vec{k}}(\vec{x}_o)
\end{equation}
where the sum runs over the allowed values of $\vec{k}$ as discussed above.
Here $\xi_{lm}^{\vec k }(\vec{x}_o)$ contains the information about the
manifold.
$T_l(k)$ is the transfer function containing the full Boltzmann physics,
e.\,g.\ the ordinary and the integrated Sachs-Wolfe effect, 
the Doppler contribution, the Silk damping and the reionization 
are taken into account.
The initial conditions are specified by $\Phi_{\vec{k}}$,
where it is assumed that they are Gaussian random fluctuations
at the early universe.
For the half-turn space $\Phi_{\vec{k}}$ has to fulfil 
the condition
\begin{equation}
\label{Eq:phi_condition}
\Phi_{-\vec k^{\,'}}^{*}\,(-1)^{n_z} \; = \; \Phi_{\vec k}
\end{equation}
where $\Phi_{\vec{k}} \in {\mathbb R}$  if $n_z=0$ and 
$\Phi_{\vec{k}} \in {\mathbb C}$ otherwise.
The assumption of initial Gaussian random fluctuations
determines the correlation of $\Phi_{\vec{k}}$ to be
\begin{equation}
\label{Eq:phi_ensemble}
\left\langle\Phi_{\vec k}^{*}\; \Phi_{\tilde {\vec k}}\right\rangle
\; = \;
P(k) \; \delta_{\vec k,\tilde {\vec k}}
\hspace{10pt} .
\end{equation}
The primordial spectrum $P(k)$ is assumed to be $P(k)\sim k^{n_s-4}$,
where $n_s$ is the spectral index. 
With the correlation (\ref{Eq:phi_ensemble})
the ensemble average $\langle ... \rangle$ of the multipole moments $C_l$
can be calculated from Eq.\,(\ref{Eq:a_lm_half_turn_space})
for a given position $\vec x_o$ of the observer.
This leads to the multipole moments $C_l$ of the half-turn space
\begin{eqnarray}
\label{Eq:Cl_ensemble_half_turn}
C_l & := &
\frac{1}{2l+1}\sum_{m=-l}^l\left\langle\left|a_{lm}\right|^2\right\rangle
\\ & = &\nonumber
\sum_{\vec k}
\frac{ T_l^2(k) \; P(k)}{2l+1} \sum_{m=-l}^l \left|Y_{lm}(\hat{k})\right|^2
\times
\\ & & \;
\left[1+(-1)^{m+n_z}\,\left(1-\delta_{0,n_x}\,\delta_{0,n_y}\right)\,
\cos\left(\left(\vec k-\vec k'\right)\cdot \vec x_o \right)\right]
\label{Eq:Cl_ensemble_half_turn_b}
\\ & = &\nonumber
\frac{1}{4\pi} \, \sum_{\vec k} T_l^2(k) \, P(k) \times
\\ & & \;
\left[1+(-1)^{n_z}\,\left(1-\delta_{0,n_x}\,\delta_{0,n_y}\right)\, \cos\left((\vec k-\vec k')\cdot \vec x_o \right)
P_l(\hat k \cdot \hat k')\right]
\label{Eq:Cl_ensemble_half_turn_c}
\hspace{4pt} .
\end{eqnarray}

The multipole moment of the half-turn space (\ref{Eq:Cl_ensemble_half_turn_c}) 
depends on the position $\vec x_o$ of the observer within the fundamental cell.
Taking the mean value of the multipole moment
(\ref{Eq:Cl_ensemble_half_turn_c})
over all observer positions leads to the simple expression
\begin{equation}
\label{Eq:Cl_ensemble_half_turn_observer}
\bar C_l^{\vec x_o} \; = \;
\frac{1}{4\pi} \, \sum_{k} T_l^2(k) \;  P(k) \; r(k)
\end{equation}
where $r(k)$ is the multiplicity of the eigenvalue $E_k=k^2$.
The multiplicity is the number of triplets $(n_x,n_y,n_z)$
which satisfy
$k = 2\pi \sqrt{\frac{n_x^2}{L_x^2}+\frac{n_y^2}{L_y^2}+\frac{n_z^2}{L_z^2}}$
and fulfil the restrictions stated at
Eqs.(\ref{Eq:eigenfunction_a}) and (\ref{Eq:eigenfunction_b}).

The ensemble average over the sky realisations of the
temperature correlation function $C(\vartheta)$ is computed by
\begin{equation}
\label{Eq:C_theta_C_l}
C(\vartheta) \; = \; \sum_l \frac{2l+1}{4\pi} \, C_l \, P_l(\cos\vartheta)
\hspace{10pt} .
\end{equation}

The above formulae allow the computation of the CMB anisotropies
when the cosmological parameters are specified.
For these we take the parameters of the $\Lambda$CDM concordance model
which are based on the WMAP 5 year data \cite{Dunkley_et_al_2009}.
The parameters are obtained from the LAMBDA website (lambda.gsfc.nasa.gov),
see the WMAP Cosmological Parameters of the model ``lcdm+sz+lens''
using the data ``wmap5+bao+snall+lyapost''.
The values are
$\Omega_{\hbox{\scriptsize bar}}=0.0474$,
$\Omega_{\hbox{\scriptsize cdm}}=0.243$,
$\Omega_{\scriptsize \Lambda}=0.709$,
and $h=0.697$ for the present day reduced Hubble constant.
The spectral index is $n_s=0.969$ and the depth to reionization
$\tau = 0.094$.
These parameters specify a flat universe
with an angular power spectrum $\delta T_l^2 = l(l+1) C_l/(2\pi)$
having its first acoustic peak at $l\simeq 220$.
The $\delta T_l^2$ spectrum is normalised to the WMAP 
best fit angular power spectrum at $l=220$ having
$\delta T_{220}^2 = 5785.6 \mu\hbox{K}^2$.


\section{The cubic half-turn space}

As described in the Introduction and in the previous section
the ensemble average of the CMB statistics depends on
the position of the observer,
but also on the sizes of the three topological lengths $L_x$, $L_y$, and $L_z$
which identify the opposing pairs of faces of the half-turn space.
In order to simplify the already complicated analysis in the first step
we devote this section to the cubic half-turn space
where all three side lengths are equal,
i.\,e.\ $L_x=L_y=L_z\equiv L$.
This allows to discuss the position dependence of
the half-turn space with respect to a single topological parameter.

To quantify the power at large angular scales by a scalar measure,
the $S(60^{\circ})$ statistic
\begin{equation}
\label{Eq:S_statistic_60}
S(60^\circ)\; := \; \int^{\cos(60^\circ)}_{-1}
d\cos\vartheta \; |C(\vartheta)|^2
\hspace{10pt}
\end{equation}
has been introduced \cite{Spergel_et_al_2003},
which measures the power in the correlation function $C(\vartheta)$
on scales larger than $60^{\circ}$.
The value of $60^{\circ}$ is arbitrary and adapted to the observed fact
that $C(\vartheta)$ almost vanishes for angles larger than this one.
Note that due to the measure $d\cos\vartheta$,
the $S(60^{\circ})$ statistic is insensitive to the behaviour
of the correlation function $C(\vartheta)$ at $\vartheta=180^\circ$.
It is sensitive for variations of $C(\vartheta)$ in the range
$60^{\circ} \lesssim \vartheta \lesssim 120^{\circ}$.

\begin{figure}
\begin{center}
\vspace*{-30pt}
\begin{minipage}{11cm}
\includegraphics[width=10.0cm]{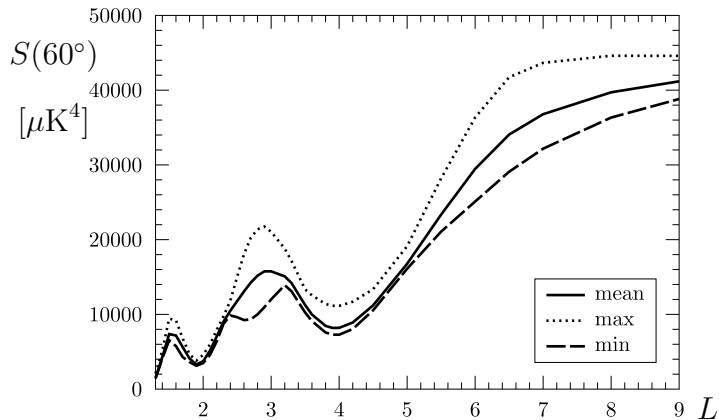}
\end{minipage}
\put(-65,-79){$L$}
\put(-325,55){$S(60^\circ)$}
\put(-322,30){[$\mu\hbox{K}^4$]}
\end{center}
\vspace*{-20pt}
\caption{\label{Fig:S60_Cubic_Half_Turn_Space}
The $S(60^\circ)$ statistic, defined in Eq.\,(\ref{Eq:S_statistic_60}),
is shown for the cubic half-turn space as a function of the
topological length scale $L$ in units of
the Hubble length $L_{\hbox{\scriptsize H}}$.
The solid curve displays the average over all positions of the observer.
The dotted and the dashed curves give the maximal and the minimal value,
respectively, of $S(60^\circ)$ in order to reveal the range of variation.
}
\end{figure}

It is important to distinguish between two different averages.
On the one hand there is the ensemble average for a single position
$\vec x_o$ of the observer
which takes the ensemble of CMB sky realisations into account.
On the other hand one can average this position dependent ensemble average
over all positions which the observer can occupy in the fundamental cell.
The position average of the ensemble averages of $S(60^\circ)$ is plotted in
figure \ref{Fig:S60_Cubic_Half_Turn_Space} as a solid curve
as a function of the side length $L$.
A variation between $2\,000 \mu\hbox{K}^4$ and $40\,000 \mu\hbox{K}^4$
depending on the side length $L$ is revealed.
Low values of power are obtained for $L$ close to the side lengths 2 and 4
in units of the Hubble length $L_{\hbox{\scriptsize H}} = c/H_0$.
In addition, the figure \ref{Fig:S60_Cubic_Half_Turn_Space}
shows the maximal (dotted curve) and the minimal (dashed curve) values
of $S(60^\circ)$ that occur among the different positions.
An asymmetric distribution can be inferred from the figure
because the difference between the mean and the maximal value is larger than
the difference between the mean and the minimal value.

Since a low value of the $S(60^\circ)$ statistic is observed
in the CMB data, the minima at the side lengths $L=1.9$ and $L=4$
in figure \ref{Fig:S60_Cubic_Half_Turn_Space} are interesting,
where values around $3500 \mu\hbox{K}^4$ and $8000 \mu\hbox{K}^4$ occur,
respectively.
These low values have to be compared with the observed ones.
We compute the correlation function $C^{\hbox{\scriptsize obs}}(\vartheta)$
from the ILC 7 year map \cite{Gold_et_al_2010} which gives 
$S_{\hbox{\scriptsize ILC}}(60^\circ) = 8\,033\,\mu\hbox{K}^4$.
By applying the KQ75 7yr mask \cite{Gold_et_al_2010} to the ILC 7 year map,
a correlation function $C^{\hbox{\scriptsize obs}}(\vartheta)$
is obtained which leads to only
$S_{\hbox{\scriptsize ILC,KQ75}}(60^\circ) = 1\,153\,\mu\hbox{K}^4$.
Note that the infinite volume concordance model has large values
which can be read off from figure \ref{Fig:S60_Cubic_Half_Turn_Space}
in the limit of large values of $L$,
i.\,e.\ at $L=9$.
It is obvious that with respect to the power on large angular scales,
the finite volume models lead to a better description of the data.

\begin{figure}
\begin{center}
\vspace*{-10pt}
\begin{minipage}{18cm}
\hspace*{-70pt}\includegraphics[width=11.0cm]{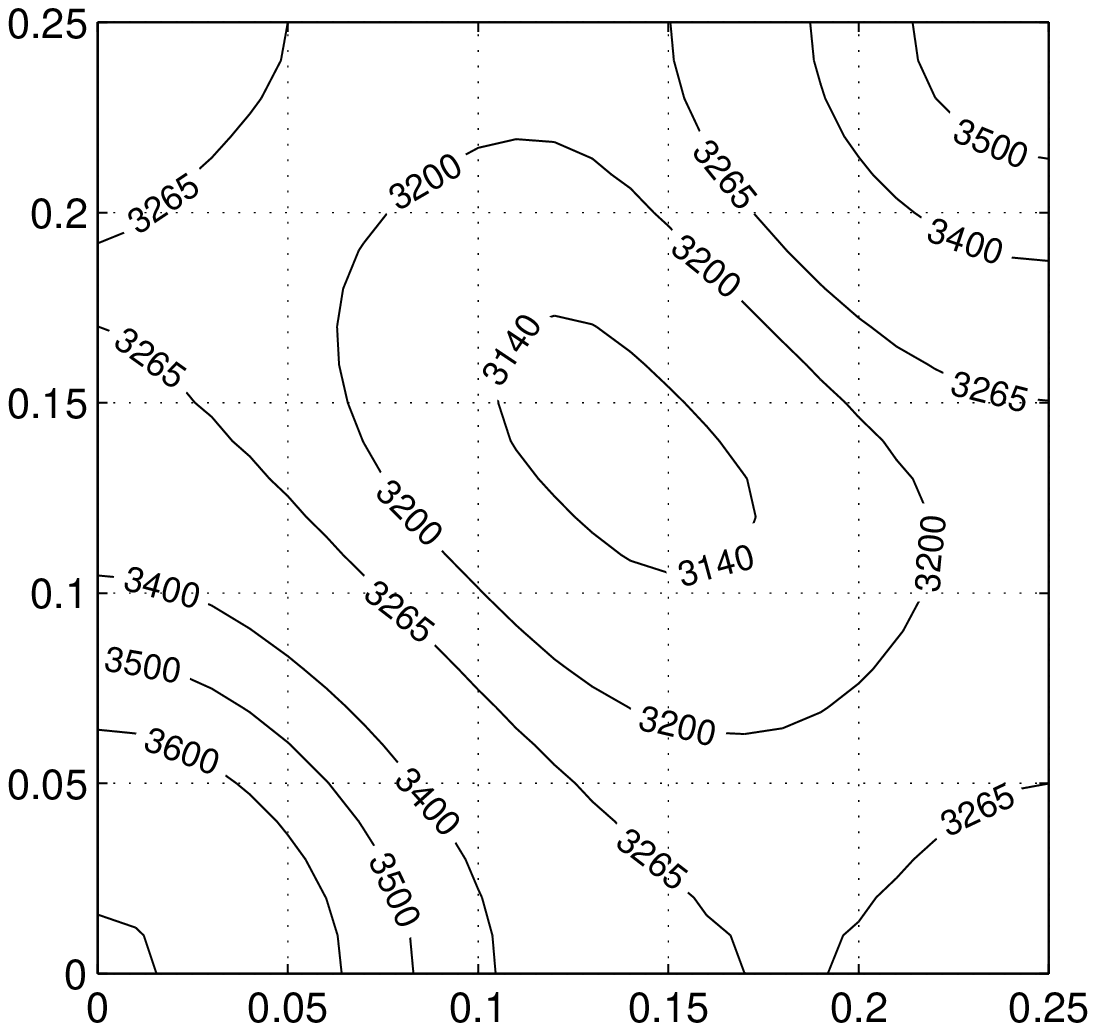}
\hspace*{-50pt}\includegraphics[width=11.0cm]{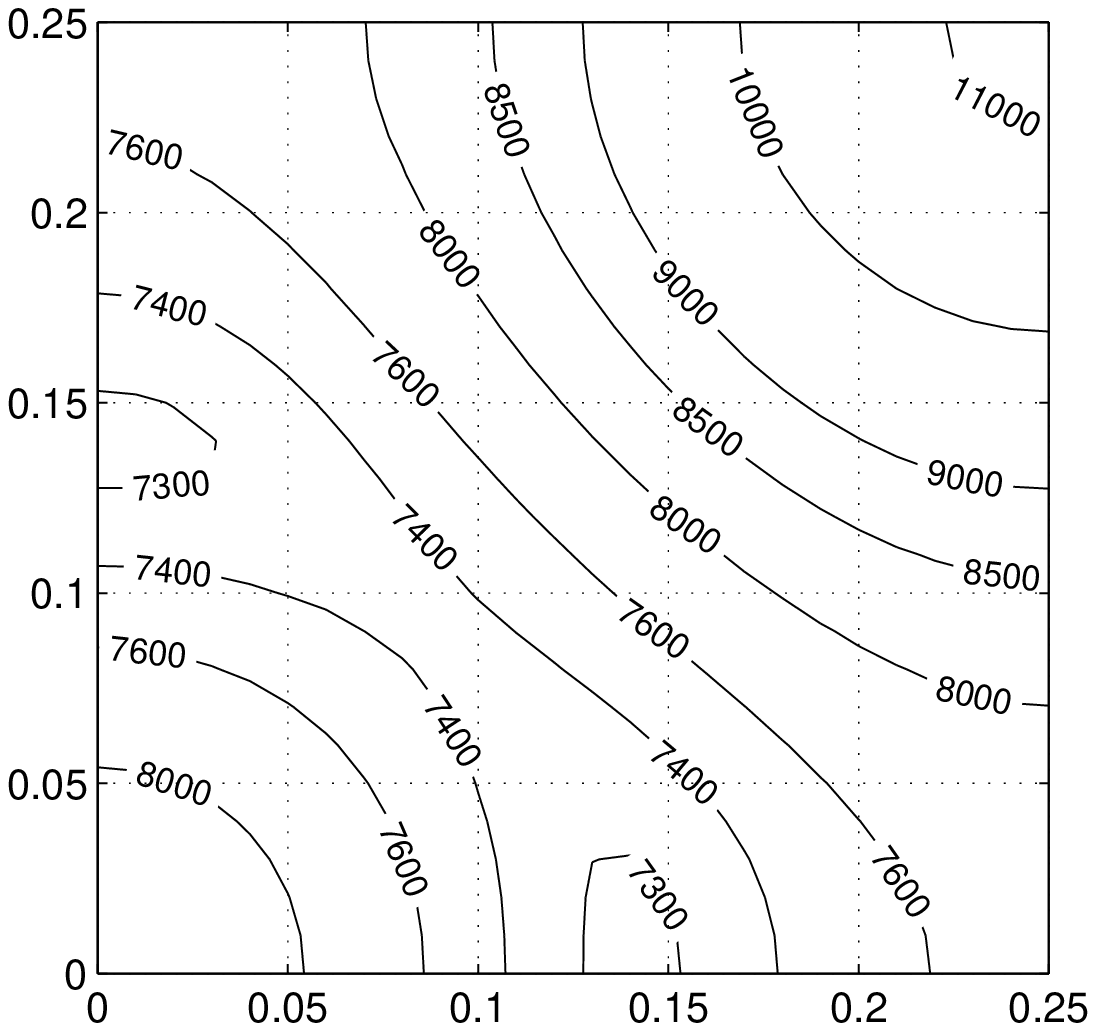}
\end{minipage}
\put(-310,-98){$x_o$}
\put(-542,84){$y_o$}
\put(-46,-98){$x_o$}
\put(-275,84){$y_o$}
\put(-475,88){$L=1.9$}
\put(-238,88){$L=4.0$}
\end{center}
\vspace*{-20pt}
\caption{\label{Fig:S60_Cubic_Half_Turn_Space_Observer}
The $S(60^\circ)$ statistic is plotted in units [$\mu\hbox{K}^4$]
in dependence on the position $(x_o,y_o)$ of the observer.
The cubic half-turn spaces with the
topological length scales $L=1.9$ (left) and $L=4$ (right) are shown.
The coordinates $(x_o,y_o)$ of these observers are given in units
of the side length $L$.
}
\end{figure}

For the two side lengths $L=1.9$ and $L=4$,
the figure \ref{Fig:S60_Cubic_Half_Turn_Space_Observer}
displays the dependence of the ensemble average of the $S(60^\circ)$
statistic on the position $(x_o,y_o)$ of the observer in the $xy$-plane.
Since there is no dependence on the $z$-coordinate,
this figure already reveals the full range of variation
in the fundamental cell.
The coordinates are, as explained in the previous section,
given in units of the side length $L$.
Due to the symmetry expressed by Eq.(\ref{Eq:Cl_ensemble_half_turn_c})
only the sixteenth part of the $x_oy_o$-plane is shown.
One can read off from figure \ref{Fig:S60_Cubic_Half_Turn_Space_Observer}
the domains where the ensemble average of the $S(60^\circ)$ statistic
drops to a minimum.
A comparison of both panels shows that the minima occur at different positions
of the observer in these two models.
As discussed below the points $(x_o,y_o)=(0,0)$ and $(0.25,0.25)$
are special points since for these positions
the correlation function $C(\vartheta)$ obtains
maximal and minimal values, respectively, at $\vartheta=180^\circ$.
These positions correspond to local maxima in
figure \ref{Fig:S60_Cubic_Half_Turn_Space_Observer}.

\begin{figure}[htb]
\vspace*{-20pt}\begin{center}
\hspace*{-40pt}
{
\begin{minipage}{20cm}
\includegraphics[width=10cm]{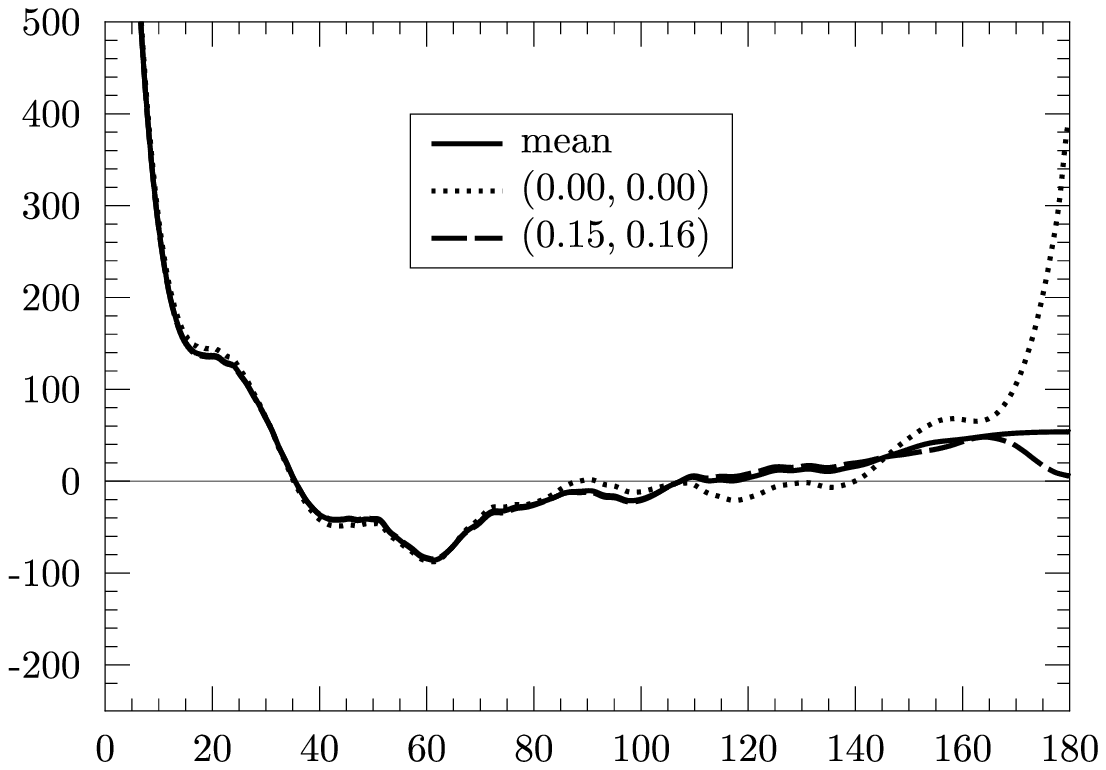}
\hspace*{-30pt}\includegraphics[width=10cm]{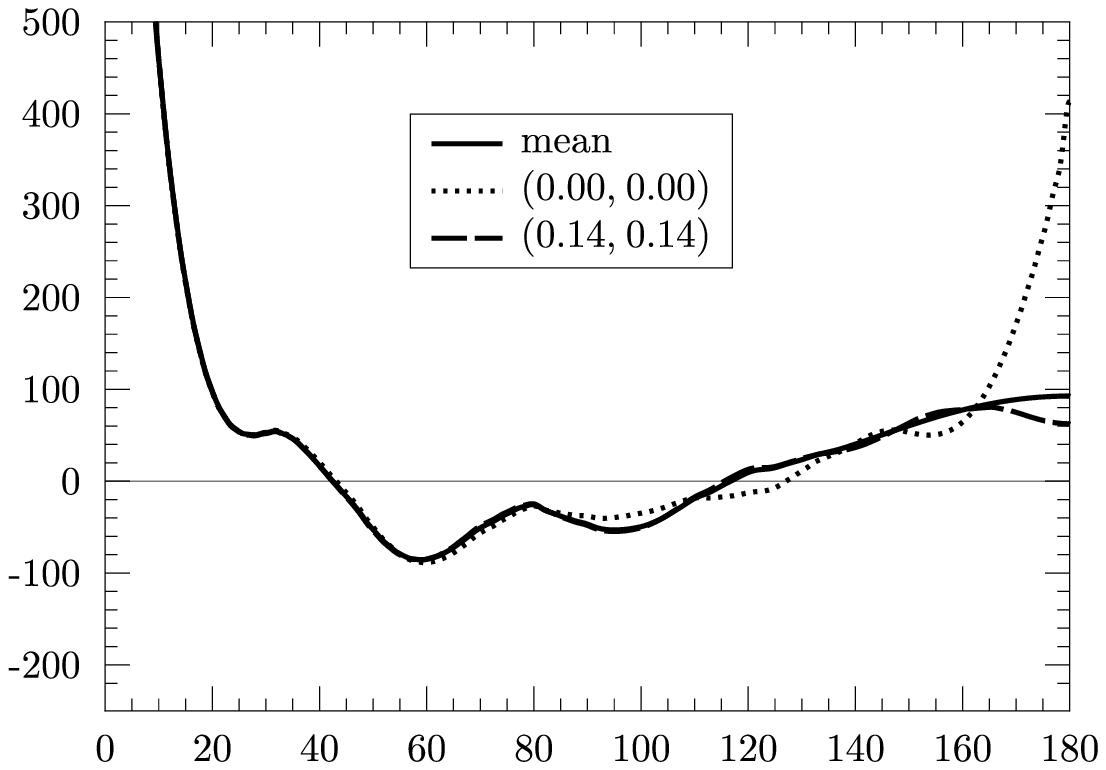}
\end{minipage}
\put(-498,55){(a)}
\put(-240,55){(b)}
\put(-569,61){$C(\vartheta)$}
\put(-571,43){[$\mu\hbox{K}^2$]}
\put(-319,-78){$\vartheta$}
\put(-310,61){$C(\vartheta)$}
\put(-312,43){[$\mu\hbox{K}^2$]}
\put(-61,-78){$\vartheta$}
\put(-380,-55){$L=1.3$}
\put(-121,-55){$L=1.9$}
\vspace*{-25pt}
}
\hspace*{-40pt}
{
\begin{minipage}{20cm}
\includegraphics[width=10cm]{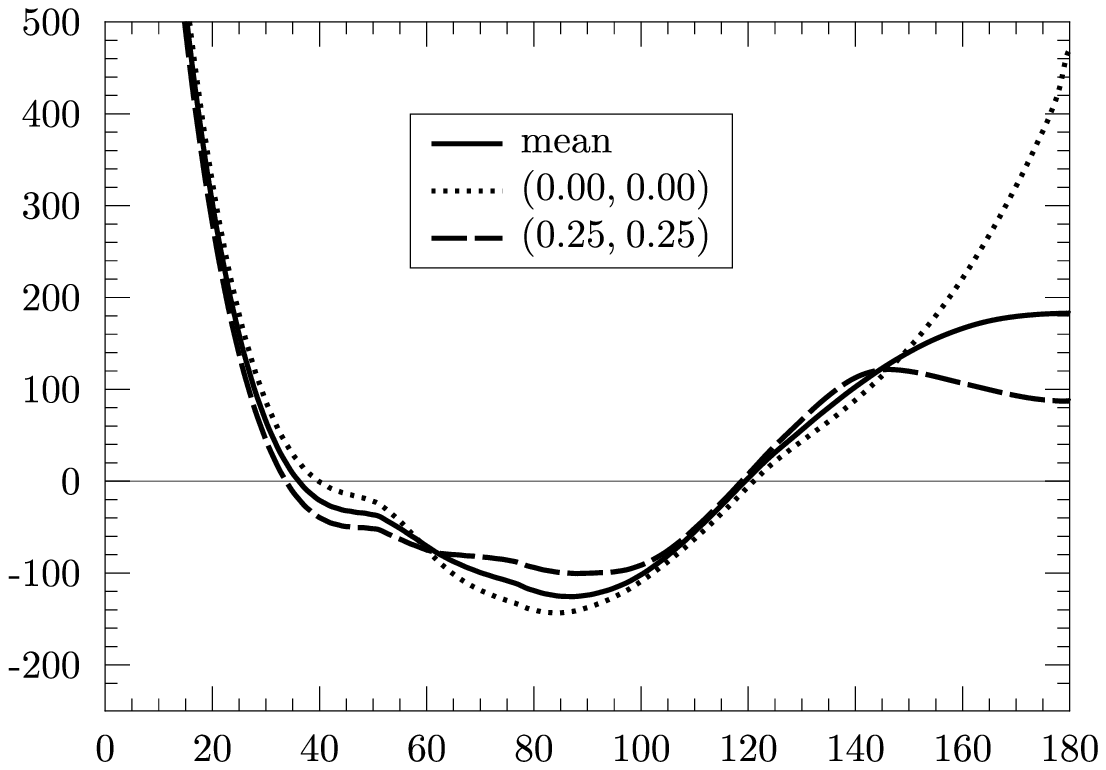}
\hspace*{-30pt}\includegraphics[width=10cm]{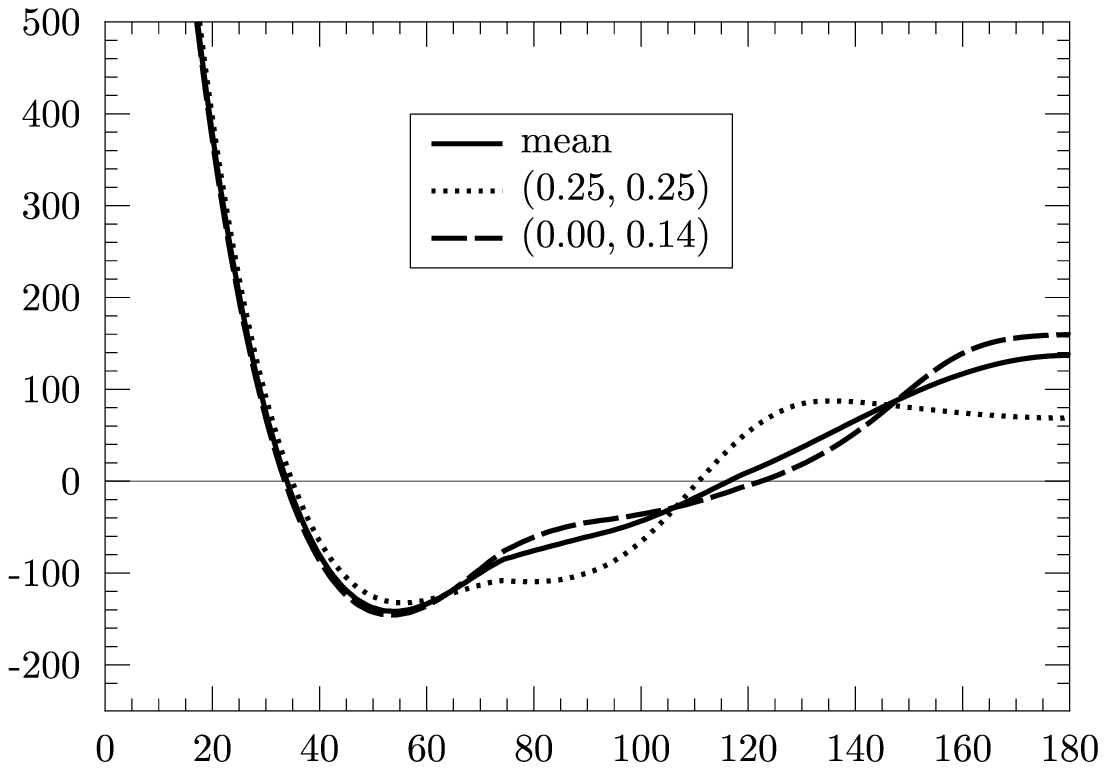}
\end{minipage}
\put(-498,55){(c)}
\put(-240,55){(d)}
\put(-569,61){$C(\vartheta)$}
\put(-571,43){[$\mu\hbox{K}^2$]}
\put(-319,-78){$\vartheta$}
\put(-310,61){$C(\vartheta)$}
\put(-312,43){[$\mu\hbox{K}^2$]}
\put(-61,-78){$\vartheta$}
\put(-380,-55){$L=2.9$}
\put(-121,-55){$L=4.0$}
\vspace*{-25pt}
}
\hspace*{-40pt}
{
\begin{minipage}{20cm}
\includegraphics[width=10cm]{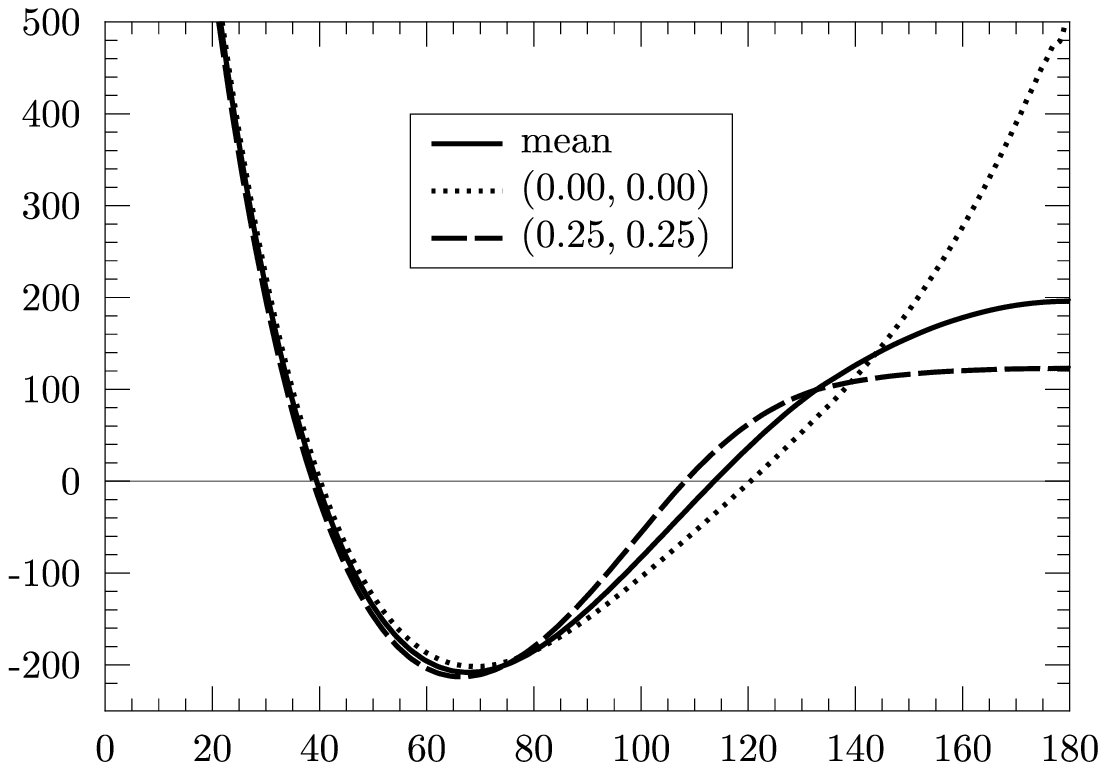}
\hspace*{-30pt}\includegraphics[width=10cm]{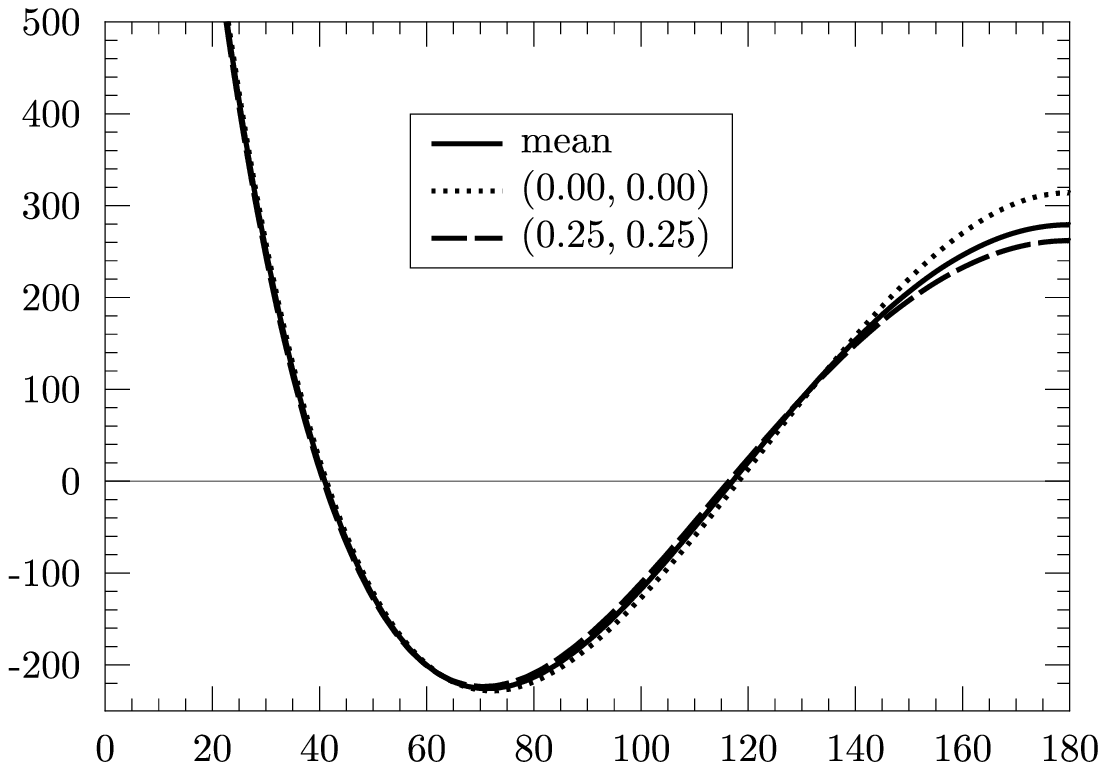}
\end{minipage}
\put(-496,55){(e)}
\put(-236,55){(f)}
\put(-569,61){$C(\vartheta)$}
\put(-571,43){[$\mu\hbox{K}^2$]}
\put(-319,-78){$\vartheta$}
\put(-310,61){$C(\vartheta)$}
\put(-312,43){[$\mu\hbox{K}^2$]}
\put(-61,-78){$\vartheta$}
\put(-380,-55){$L=6.0$}
\put(-121,-55){$L=9.0$}
\vspace*{-15pt}
}
\end{center}
\caption{
\label{Fig:Correlation_Cubic_Half_Turn_Space}
The temperature correlation $C(\vartheta)$
is shown for the cubic half-turn space for the six
topological lengths $L = 1.3$, 1.9, 2.9, 4.0, 6.0, and 9.0.
The average over all positions of the observer is plotted as
a solid curve.
The dashed curve belongs to the position with the smallest value of
$S(60^\circ)$ and the dotted one to the largest value of $S(60^\circ)$.
}
\vspace*{-10pt}
\end{figure}

Since the $S(60^\circ)$ statistic integrates the correlation function
$C(\vartheta)$ no information about the angular dependence $\vartheta$
is preserved.
Thus the figure \ref{Fig:Correlation_Cubic_Half_Turn_Space}
displays for six different topological scales $L$ the
correlation function $C(\vartheta)$.
The solid curve shows the position average of the ensemble average
of $C(\vartheta)$
whereas the dotted and dashed curves show the correlation functions
for the positions belonging to the extremal values of $S(60^\circ)$.

\begin{figure}[htb]
\vspace*{-20pt}\begin{center}
\hspace*{-40pt}
{
\begin{minipage}{20cm}
\includegraphics[width=10cm]{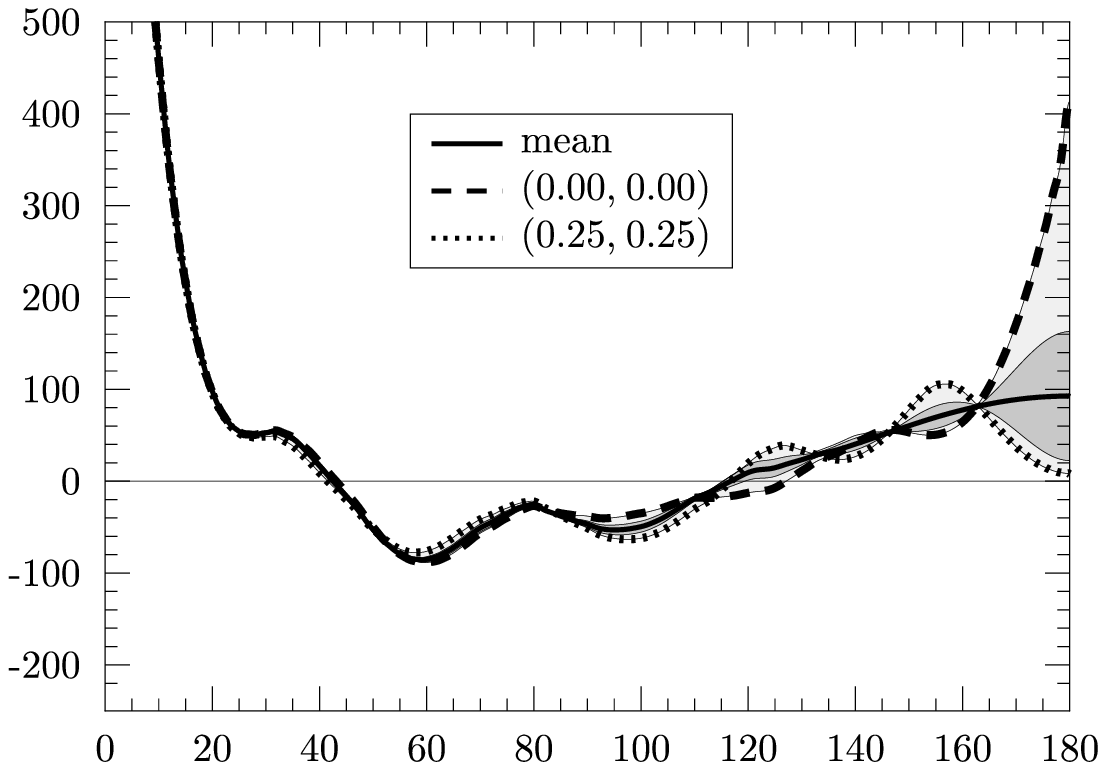}
\hspace*{-30pt}\includegraphics[width=10cm]{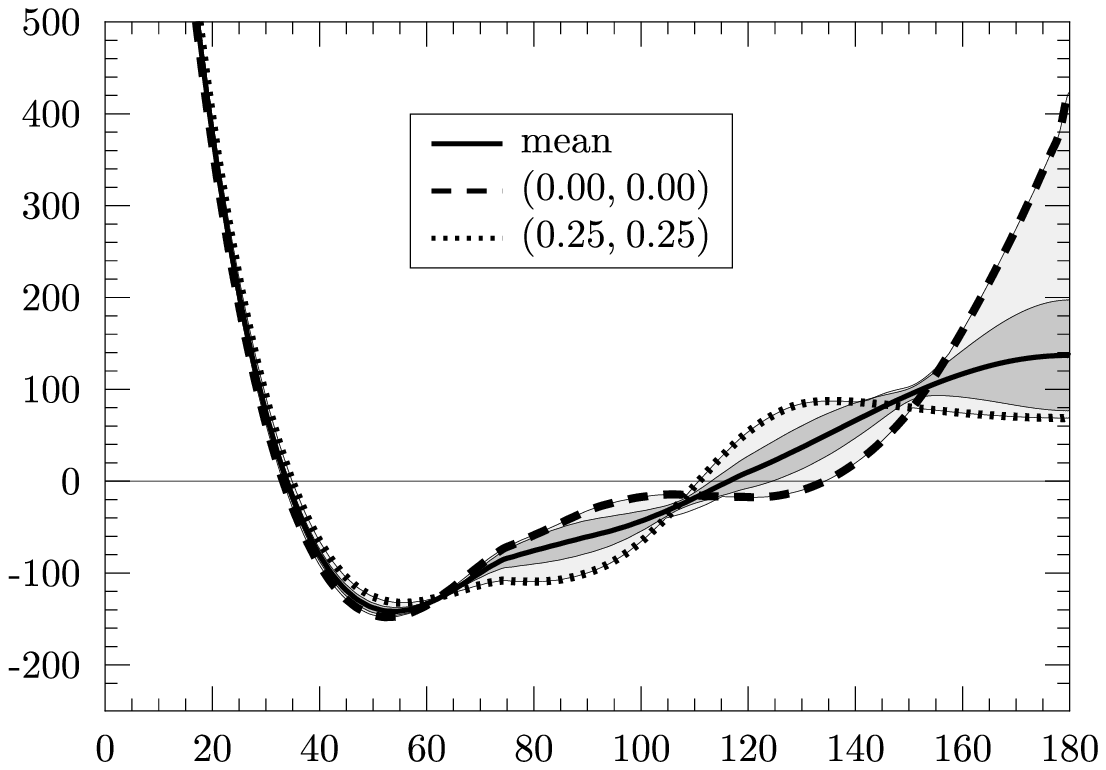}
\end{minipage}
\put(-498,55){(a)}
\put(-240,55){(b)}
\put(-569,61){$C(\vartheta)$}
\put(-571,43){[$\mu\hbox{K}^2$]}
\put(-319,-78){$\vartheta$}
\put(-310,61){$C(\vartheta)$}
\put(-312,43){[$\mu\hbox{K}^2$]}
\put(-61,-78){$\vartheta$}
\put(-380,-55){$L=1.9$}
\put(-121,-55){$L=4.0$}
\vspace*{-25pt}
}
\end{center}
\caption{
\label{Fig:Correlation_Cubic_Half_Turn_Space_mean_stand_max_min}
The ensemble average of the temperature correlation function $C(\vartheta)$
is shown for the cubic half-turn space for the
topological lengths $L = 1.9$ and 4.0.
The average over all positions of the observer is plotted as
a solid curve and its standard deviation as a dark grey band.
The distribution of the correlation function $C(\vartheta)$
depending on the position of the observer is given as a light grey band.
The correlation functions for two observers 
having extreme values in $C(\vartheta)$ at $\vartheta=180^\circ$ are plotted.
In the legend the coordinates $(x_o,y_o)$ of these observers are given in units
of the side length $L$.
}
\vspace*{-10pt}
\end{figure}

Above we discussed the dependence on the position of the observer
in the case of the $S(60^\circ)$ statistic for two models,
see figure \ref{Fig:S60_Cubic_Half_Turn_Space_Observer}.
There we already point out that the correlation function $C(\vartheta)$
obtains at $\vartheta=180^\circ$ extremal values for two
special positions which in turn lead to a local maximum
with respect to the $S(60^\circ)$ statistic.
The maximal value of $C(180^\circ)$ always occurs at the point
$(x_o,y_o)=(0,0)$,
whereas at $(x_o,y_o)=(0.25,0.25)$ it drops to a minimum.
Figure \ref{Fig:Correlation_Cubic_Half_Turn_Space_mean_stand_max_min}
demonstrates this observational fact for the two models with
side length $L=1.9$ (left) and $4.0$ (right).
That these positions of the observer are special is revealed
by Eq.\,(\ref{Eq:Cl_ensemble_half_turn_c})
which simplifies for the above two observer positions.
When the argument of the cosine is written explicitly as
$$
\cos\left( (\vec k - \vec k\,')\, \vec x_o \right) \; = \;
\cos\left( \pi \, ( 4 n_x x_o + 4 n_y y_o ) \right)
\hspace{10pt} ,
$$
one obtains 1 for $(x_o,y_o)=(0,0)$ and
$(-1)^{n_x+n_y}$ for $(x_o,y_o)=(0.25,0.25)$.
These are the extreme situations which can occur with respect
to the cancellation of neighbouring terms in the sum.
Eq.\,(\ref{Eq:Cl_ensemble_half_turn_c}) reduces
for $(x_o,y_o)=(0,0)$ to
\begin{equation}
\label{Eq:C_l_0_0}
C_l \, = \,
\sum_{\vec k} \frac{T_l^2(k) \, P(k)}{4\pi} \,
\left[1+(-1)^{n_z}\,\left(1-\delta_{0,n_x}\,\delta_{0,n_y}\right) \,
P_l(\hat k \cdot \hat k')\right]
\end{equation}
and for $(x_o,y_o)=(0.25,0.25)$ to
\begin{equation}
\label{Eq:C_l_0.25_0.25}
C_l \, = \,
\sum_{\vec k} \frac{T_l^2(k) \, P(k)}{4\pi} \,
\left[1+(-1)^{n_x+n_y+n_z}\,\left(1-\delta_{0,n_x}\,\delta_{0,n_y}\right) \,
P_l(\hat k \cdot \hat k')\right]
\hspace{2pt} .
\end{equation}
The sum over $\vec k$ runs over the integers $n_x$, $n_y$, and $n_z$.
The complicated structure of the transfer function $T_l(k)$
and the presence of the Legendre function $P_l(\hat k \cdot \hat k')$
prevent the derivation of analytical expressions
which would show how these values of $C_l$ lead to extremal values for
$C(180^\circ)$.
Note that Eq.\,(\ref{Eq:C_theta_C_l})
for the computation of $C(\vartheta)$ reduces for $\vartheta=180^\circ$ to
\begin{equation}
\label{Eq:C_180_C_l}
C(180^\circ) \; = \; \sum_l (-1)^l \; \frac{2l+1}{4\pi} \, C_l
\hspace{10pt} .
\end{equation}
It turns out that the argument of the Legendre function
$$
\hat k \cdot \hat k' \; = \;
\frac{-n_x^2-n_y^2+(n_z/2)^2}{n_x^2+n_y^2+(n_z/2)^2}
$$
is for most summands close to $\hat k \cdot \hat k' \simeq -1$,
since the terms with either $n_x\gg n_z$ or $n_y\gg n_z$ or both
dominate those terms with $n_z\gg \max(n_x,n_y)$.
Thus for most terms one approximately gets
$P_l(\hat k \cdot \hat k')=(-1)^l$.
Furthermore, the absence of $n_x$ and $n_y$ in
the sign factor $(-1)^{n_z}$ in Eq.\,(\ref{Eq:C_l_0_0})
causes the coherent addition of all terms with the same $n_z$
but different $n_x$ and $n_y$.
The reverse situation is realised in Eq.\,(\ref{Eq:C_l_0.25_0.25}).
The numeric reveals that Eq.\,(\ref{Eq:C_l_0_0}) leads to extreme
fluctuations in $C_l$ alternating in $l$,
where even values of $l$ yield large $C_l$'s and odd $l$'s small $C_l$'s.
The factor $(-1)^l$ in Eq.\,(\ref{Eq:C_180_C_l}) leads then to a maximal
value of $C(180^\circ)$.
Although the fluctuations of $C_l$ in Eq.\,(\ref{Eq:C_l_0.25_0.25})
are less pronounced than in Eq.\,(\ref{Eq:C_l_0_0}),
the crucial difference is that now odd values of $l$ belong to the
large values of $C_l$ (for not too large values of $l$)
which in turn leads to a small value of $C(180^\circ)$.
This discussion highlights
that inhomogeneous spaces have much more freedom than homogeneous spaces
with respect to their CMB statistics.


\section{General half-turn spaces}

In the previous section only half-turn spaces are considered
where all three topological lengths $L_x$, $L_y$, and $L_z$ are identical.
This restriction to cubic half-turn spaces is now dropped.
The analysis of the last section has shown
that cubic models with a topological length $L\simeq 4$ yield especially
low values for the $S(60^\circ)$ statistic.
This corresponds to models with a volume ${\cal V} = L^3$,
where this volume is specified in units of the Hubble volume
$L_{\hbox{\scriptsize H}}^3$.
The Hubble length $L_{\hbox{\scriptsize H}} = c/H_0$ is close to
$L_{\hbox{\scriptsize H}} \simeq 4.28\,\hbox{Gpc}$ for $h\simeq 0.7$
leading to a physical volume of
${\cal V}_{\hbox{\scriptsize phys}} \simeq 5\,000\,\hbox{Gpc}^3$.
This is the same volume as that of the cubic torus,
i.\,e.\ a homogeneous space form,
which gives a good description
\cite{Aurich_Janzer_Lustig_Steiner_2007} of the WMAP data.
In order to obtain a volume which does not depend on
the Hubble constant $H_0$, one can consider the ratio
${\cal V}_{\hbox{\scriptsize phys}}/{\cal V}_{\hbox{\scriptsize sls}}$,
where ${\cal V}_{\hbox{\scriptsize sls}}$ is the volume within the
surface of last scattering.
For the cubic half-turn space as well as for the cubic torus, one obtains
${\cal V}_{\hbox{\scriptsize phys}}/{\cal V}_{\hbox{\scriptsize sls}} \simeq 0.42$.
It is worthwhile to note
that also in the case of the three spherical space forms
studied in \cite{Aurich_Lustig_Steiner_2005a} similar volumes are
found which provide a good match with the WMAP data.
For the dodecahedral space, the binary octahedral space,
and the binary tetrahedral space, one finds
${\cal V}_{\hbox{\scriptsize phys}}/{\cal V}_{\hbox{\scriptsize sls}} \simeq 0.47$,
$0.40$, and $0.37$, respectively \cite{Aurich_Lustig_Steiner_2005a}.
Thus it is natural to compare half-turn spaces
where the volume ${\cal V} = L^3$ is hold fixed
by using the parameterisation
\begin{equation}
\label{Eq:lengths}
L_x = \alpha L
\hspace{10pt} , \hspace{10pt} 
L_y = \beta L
\hspace{10pt} , \hspace{10pt} 
L_z = \frac{L}{\alpha \beta}
\hspace{10pt} .
\end{equation}
This provides for $L=4$ a parametric plane spanned by $\alpha$ and $\beta$
which is still too large for a systematic numerical search.
We confine here to two lines in the $\alpha\beta$-plane.
The first line is obtained by setting $\beta=1$,
and the second line is the ``diagonal'' in the $\alpha\beta$-plane
by setting $\alpha=\beta$.

\begin{figure}
\begin{center}
\vspace*{-30pt}
\begin{minipage}{11cm}
\includegraphics[width=10.0cm]{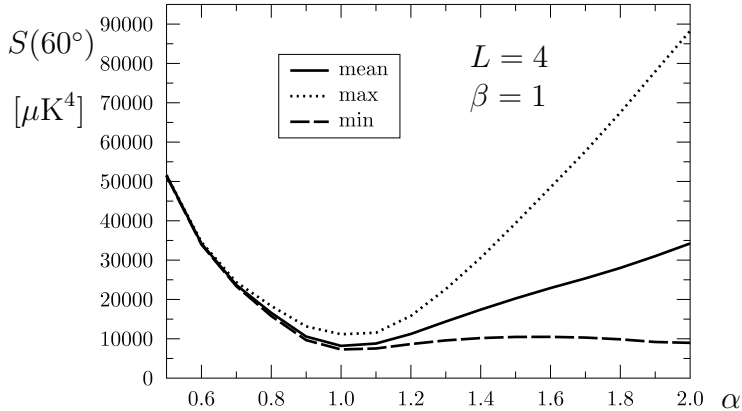}
\end{minipage}
\put(-60,-79){$\alpha$}
\put(-330,55){$S(60^\circ)$}
\put(-328,30){[$\mu\hbox{K}^4$]}
\put(-155,50){$L=4$}
\put(-155,35){$\beta=1$}
\end{center}
\vspace*{-20pt}
\caption{\label{Fig:S60_General_Half_Turn_Space}
The $S(60^\circ)$ statistic, defined in Eq.\,(\ref{Eq:S_statistic_60}),
is shown for general half-turn spaces with volume ${\cal V} = 64$
as a function of the distortion parameter $\alpha$.
The parameter $\beta$ is fixed as $\beta=1$, i.\,e.\ $L_y=4$.
The solid curve displays the average over all positions of the observer.
The dotted and the dashed curves give the maximal and the minimal value,
respectively, of $S(60^\circ)$.
}
\end{figure}

\begin{figure}
\begin{center}
\vspace*{-30pt}
\begin{minipage}{11cm}
\includegraphics[width=10.0cm]{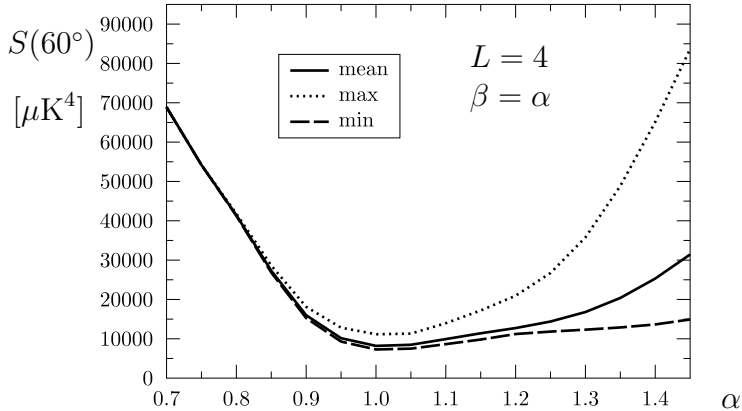}
\end{minipage}
\put(-60,-79){$\alpha$}
\put(-330,55){$S(60^\circ)$}
\put(-328,30){[$\mu\hbox{K}^4$]}
\put(-155,50){$L=4$}
\put(-155,35){$\beta=\alpha$}
\end{center}
\vspace*{-20pt}
\caption{\label{Fig:S60_General_Half_Turn_Space_diagonal}
The $S(60^\circ)$ statistic
is shown for general half-turn spaces with volume ${\cal V} = 64$
as a function of the distortion parameter $\alpha$.
The parameter $\beta$ is specified as $\beta=\alpha$.
The solid curve displays the average over all positions of the observer.
The dotted and the dashed curves give the maximal and the minimal value,
respectively, of $S(60^\circ)$.
}
\end{figure}

The figures \ref{Fig:S60_General_Half_Turn_Space}
and \ref{Fig:S60_General_Half_Turn_Space_diagonal}
show the $S(60^\circ)$ statistic for these two parametric curves.
The $S(60^\circ)$ statistic is based on the correlation function
$C(\vartheta)$ computed from Eq.\,(\ref{Eq:C_theta_C_l})
which takes the ensemble average of sky realisations into account.
The solid curves display the average over all positions of the observer.
In order to reveal the range of variation with respect to the observer position,
these figures also show the maximal and the minimal values
of the $S(60^\circ)$ statistic that occur among the various positions.
One observes that for small values of $\alpha$,
the range of variation diminishes.
This can be understood as follows.
The inhomogeneity is due to the transformation in the $z$-direction
which involves the rotation by $\pi$.
A necessary requirement for the observability of inhomogeneity is
that the diameter $D_{\hbox{\scriptsize sls}}$ of the surface of last
scattering is smaller than the topological length scale
$L_z = L/(\alpha\beta)$.
The $z$-transformation is observable for
$\alpha\beta \geq L/D_{\hbox{\scriptsize sls}}$.
The set of cosmological parameters of the concordance model
used in this paper leads to a diameter $D_{\hbox{\scriptsize sls}}=6.44$.
The transition takes place for the case $\beta=1$
shown in figure \ref{Fig:S60_General_Half_Turn_Space}
at $\alpha\simeq 0.62$,
and for the other case $\beta=\alpha$
shown in figure \ref{Fig:S60_General_Half_Turn_Space_diagonal}
at $\alpha\simeq 0.79$.
It is striking to see that the variability with respect to the
observer position sets in at exactly these values of $\alpha$.
For smaller values of $\alpha$ the topology mimics that of
the slab space which is, however, homogeneous
\cite{Cresswell_Liddle_Mukherjee_Riazuelo_2006}.
In addition, for $\alpha>1$ there are always
positions for which the $S(60^\circ)$ statistic has nearly as
small values as for the cubic case $\alpha=1$,
although there are positions for which values almost as large as
90\,000\,$\mu\hbox{K}^4$ occur (at $\alpha=2$).

\begin{table}
\centering
\begin{tabular}{c|c|c|c}
$\alpha$&$\bar S^{\vec x_o}(60^\circ)$&$\hbox{min}_{\vec x_o}(S(60^\circ))$
&$\hbox{max}_{\vec x_o}(S(60^\circ))$\\
\hline
0.5 & 51494 & 51371 & 51628\\
0.7 & 23428 & 23358 & 24377\\
1.0 &  7769 &  7282 & 11150\\
1.4 & 16188 & 10185 & 30576\\
2.0 & 29566 &  8971 & 88285\\
\end{tabular}
\caption{\label{S60_Values_general_Half_Turn_Space}
For five values of $\alpha$
the values of the $S(60^\circ)$ statistic are given
in units [$\mu\hbox{K}^4$] which are shown as the three curves
in figure \ref{Fig:S60_General_Half_Turn_Space}
($L=4$ and $\beta=1$),
i.\,e.\ the mean value as well as the two extrema.
}
\end{table}

\begin{figure}
\begin{center}
\vspace*{-10pt}
{
\begin{minipage}{18cm}
\hspace*{-70pt}\includegraphics[width=11.0cm]{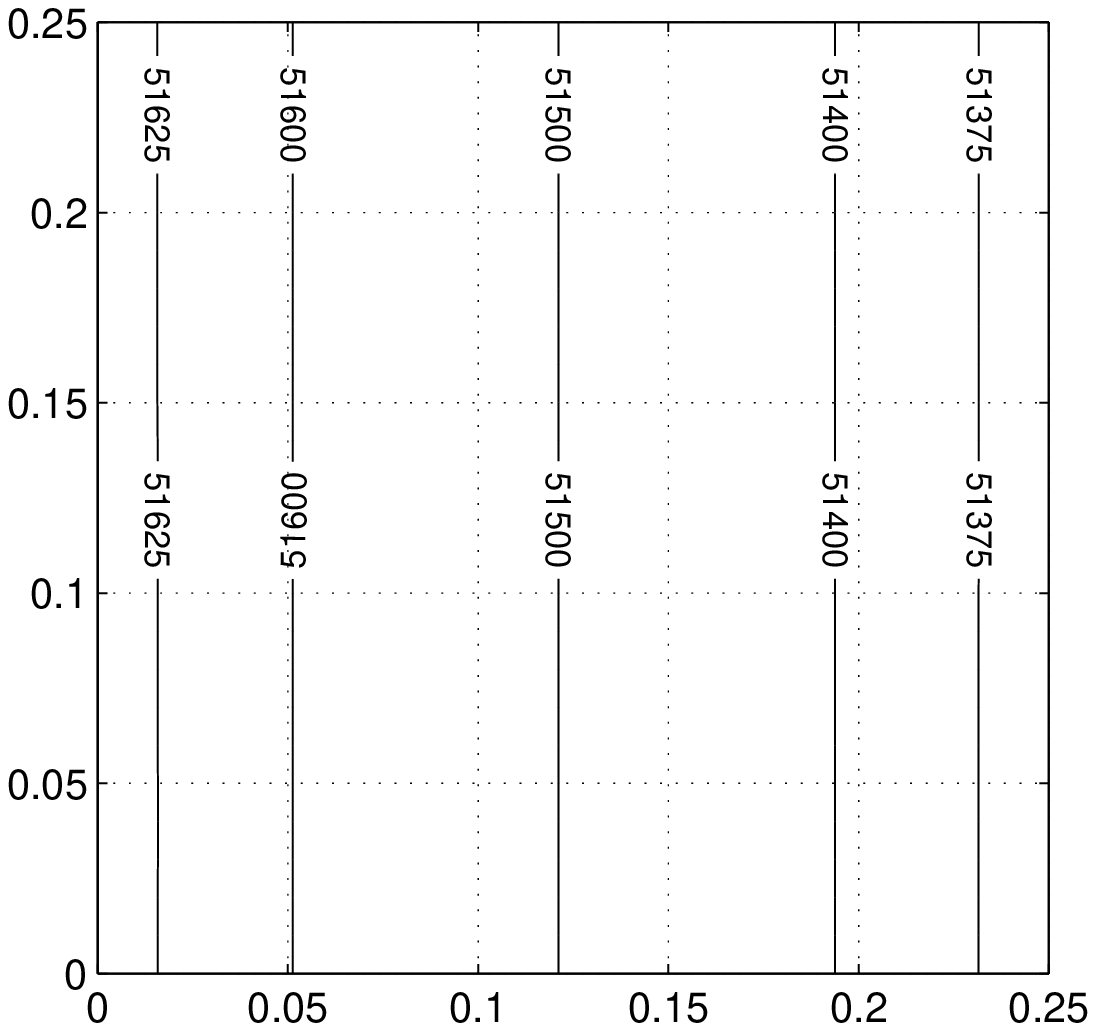}
\hspace*{-50pt}\includegraphics[width=11.0cm]{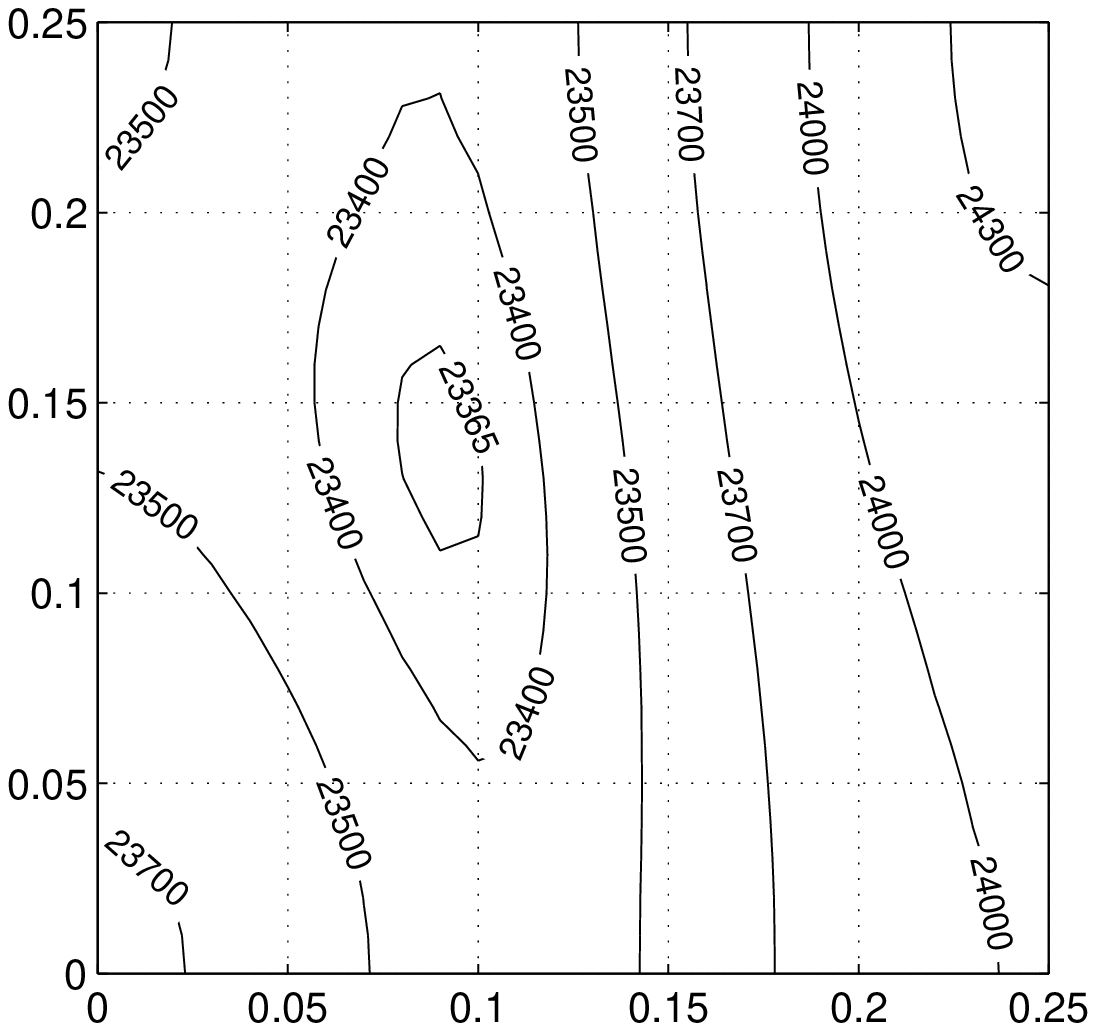}
\end{minipage}
\put(-310,-98){$x_o$}
\put(-542,84){$y_o$}
\put(-46,-98){$x_o$}
\put(-275,84){$y_o$}
\put(-470,50){$\alpha=0.5$}
\put(-225,88){$\alpha=0.7$}
}
{
\begin{minipage}{18cm}
\hspace*{-70pt}\includegraphics[width=11.0cm]{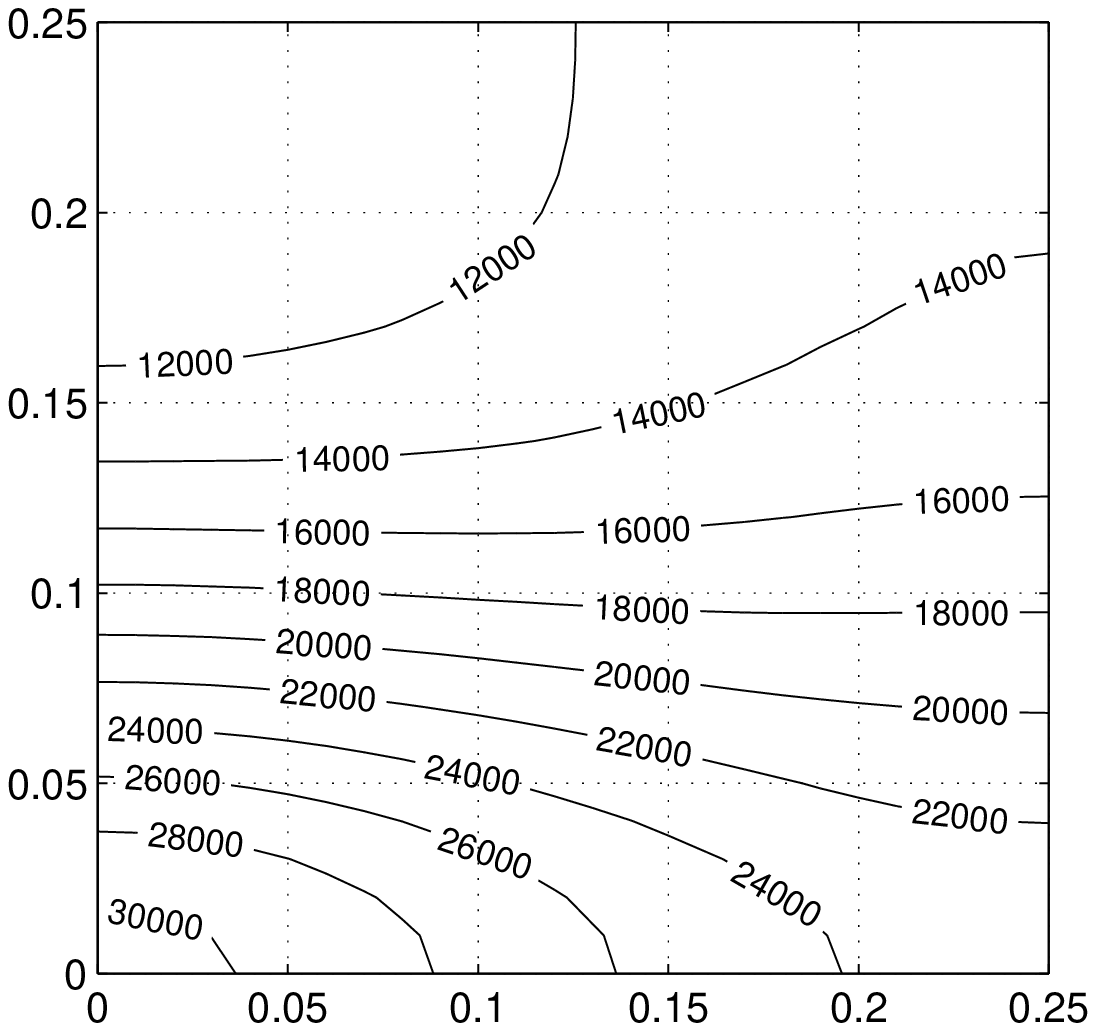}
\hspace*{-50pt}\includegraphics[width=11.0cm]{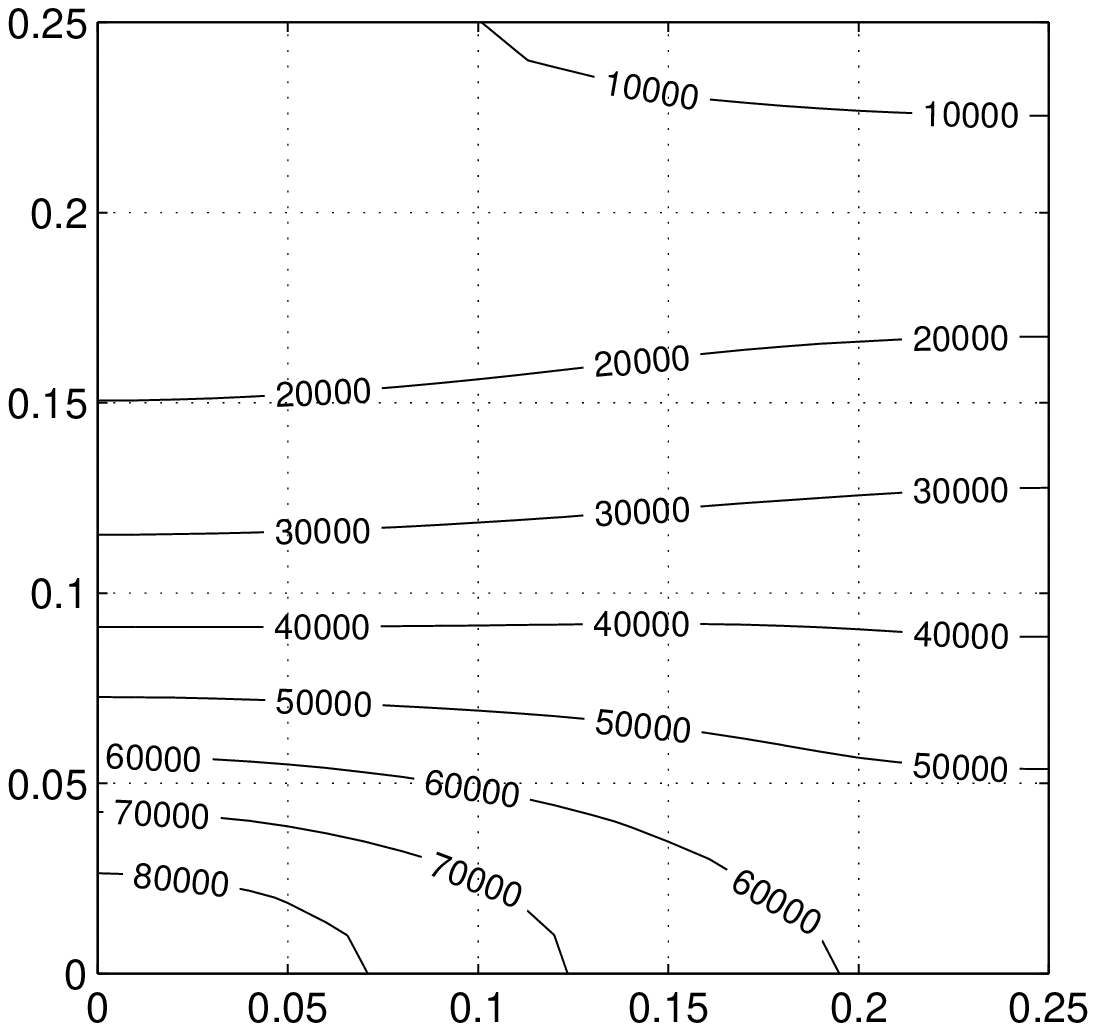}
\end{minipage}
\put(-310,-98){$x_o$}
\put(-542,84){$y_o$}
\put(-46,-98){$x_o$}
\put(-275,84){$y_o$}
\put(-500,82){$\alpha=1.4$}
\put(-235,82){$\alpha=2.0$}
}
\end{center}
\vspace*{-20pt}
\caption{\label{Fig:S60_General_Half_Turn_Space_Observer}
The $S(60^\circ)$ statistic is plotted in units [$\mu\hbox{K}^4$]
in dependence on the position $(x_o,y_o)$ of the observer.
The generic half-turn spaces with the distortion parameters
$\alpha=0.5$, $\alpha=0.7$, $\alpha=1.4$, and $\alpha=2.0$ are shown.
The parameter $\beta$ is fixed as $\beta=1$.
The coordinates $(x_o,y_o)$ of these observers are given in units
of the side lengths $L_x$ and $L_y$.
}
\end{figure}

We now discuss the case $\beta=1$ in more detail.
The figure \ref{Fig:S60_General_Half_Turn_Space_Observer}
shows the dependence of the $S(60^\circ)$ statistic on the observer position
$(x_o,y_o)$ for four selected values of $\alpha$.
The cubic case $\alpha=1$ is already shown in
figure \ref{Fig:S60_Cubic_Half_Turn_Space_Observer}.
The panel \ref{Fig:S60_General_Half_Turn_Space_Observer}(a) shows
the case with very little variability belonging to $\alpha=0.5$
which is below the critical value $\alpha\simeq 0.62$.
Here, the $S(60^\circ)$ statistic varies only marginally between
51\,371\,$\mu\hbox{K}^4$ and 51\,628\,$\mu\hbox{K}^4$,
see table \ref{S60_Values_general_Half_Turn_Space}.
This variability increases with increasing value of $\alpha$ as is revealed
by the next panels and by table \ref{S60_Values_general_Half_Turn_Space}.
For $\alpha\geq 0.7$ the maximal values again occur at the special points
$(x_o,y_o)=(0,0)$ or $(x_o,y_o)=(0.25,0.25)$.
In the case of the more interesting position belonging to the minimum of
the $S(60^\circ)$ statistic,
there are no such distinguished positions.
For values of $\alpha$ larger than one, the strongest variation takes place
with respect to the coordinate $y_o$
as revealed by the more or less horizontal lines.
The symmetric case $\alpha=1$ possesses diagonal lines in the $x_oy_o$-plane
as shown in figure \ref{Fig:S60_Cubic_Half_Turn_Space_Observer}.

\begin{figure}[htb]
\vspace*{-20pt}\begin{center}
\hspace*{-40pt}
{
\begin{minipage}{20cm}
\includegraphics[width=10cm]{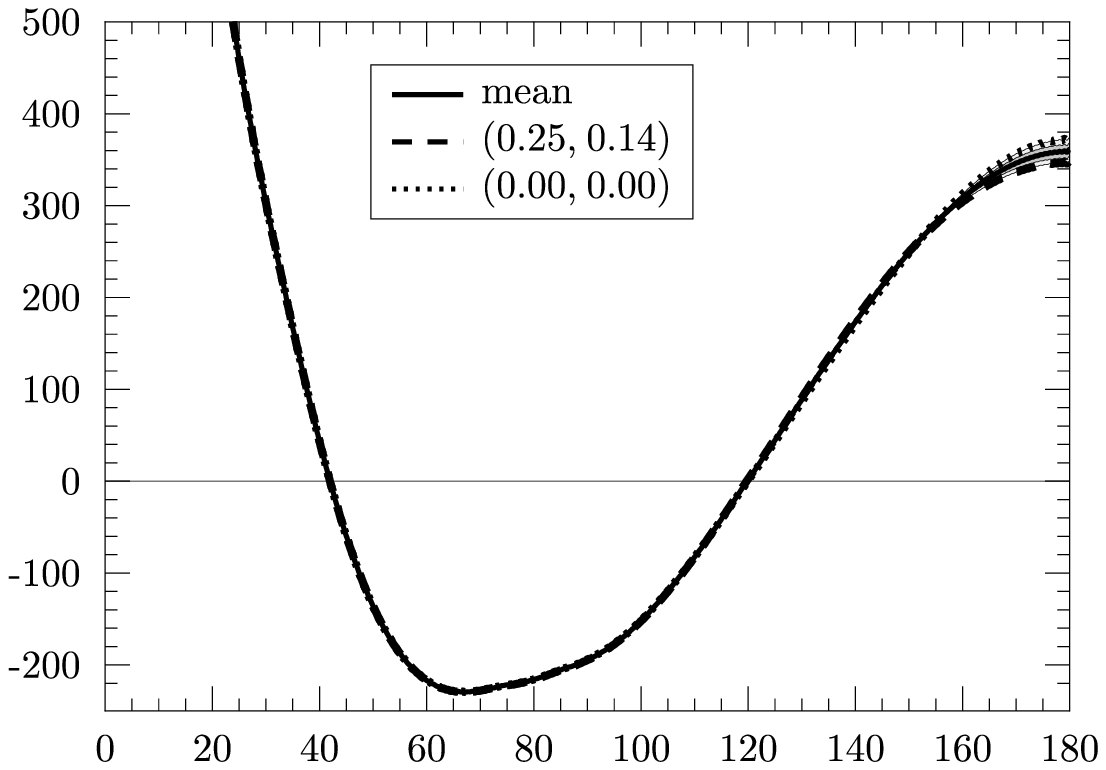}
\hspace*{-30pt}\includegraphics[width=10cm]{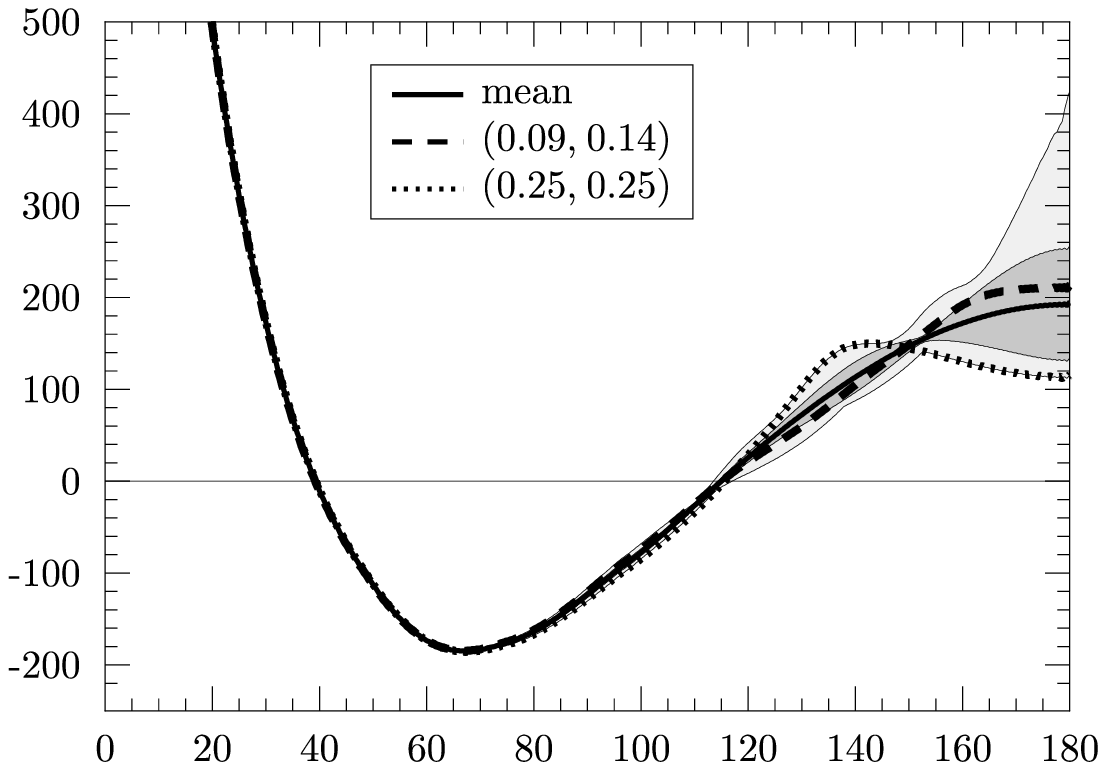}
\end{minipage}
\put(-495,55){(a)}
\put(-236,55){(b)}
\put(-569,61){$C(\vartheta)$}
\put(-571,43){[$\mu\hbox{K}^2$]}
\put(-319,-78){$\vartheta$}
\put(-310,61){$C(\vartheta)$}
\put(-312,43){[$\mu\hbox{K}^2$]}
\put(-61,-78){$\vartheta$}
\put(-380,-40){$\alpha=0.5$}
\put(-380,-55){$\beta=1.0$}
\put(-121,-40){$\alpha=0.7$}
\put(-121,-55){$\beta=1.0$}
\vspace*{-25pt}
}
\hspace*{-40pt}
{
\begin{minipage}{20cm}
\includegraphics[width=10cm]{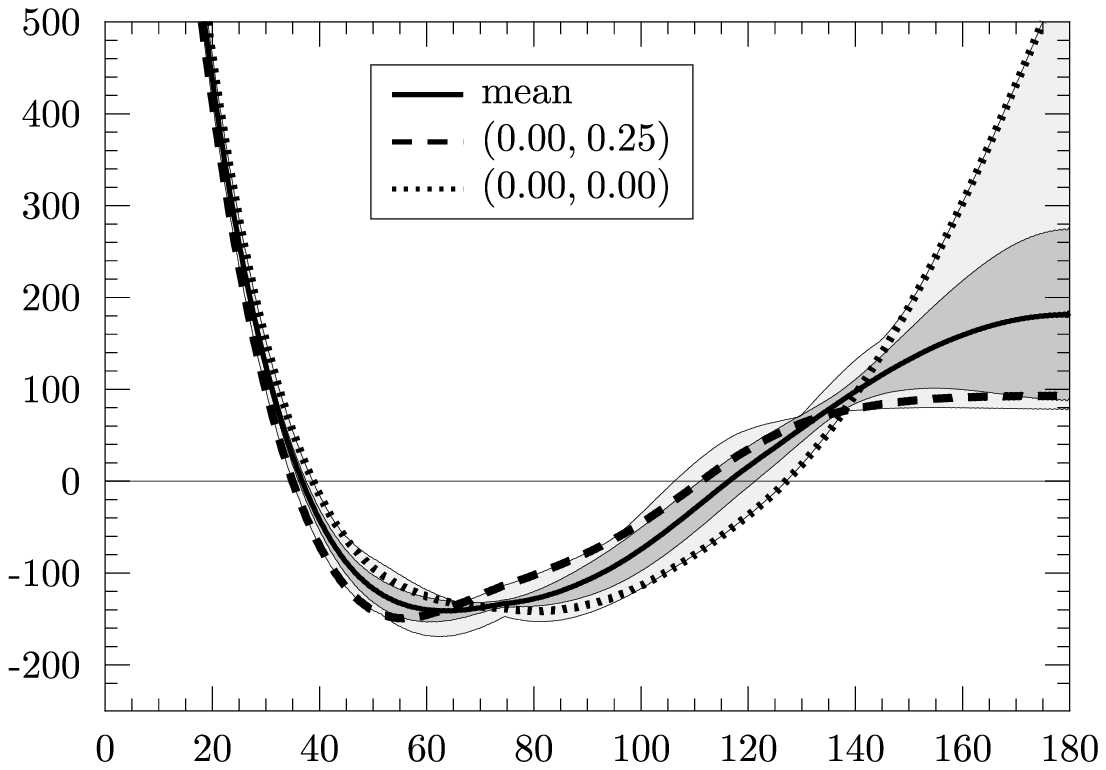}
\hspace*{-30pt}\includegraphics[width=10cm]{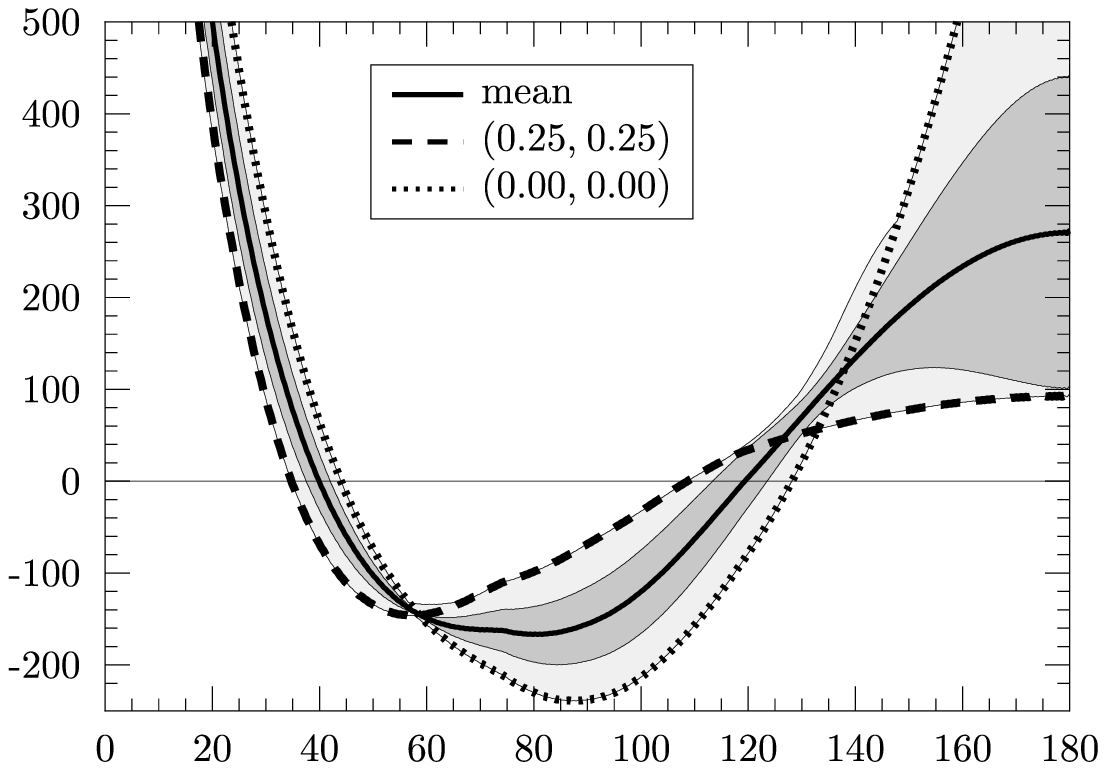}
\end{minipage}
\put(-495,55){(c)}
\put(-236,55){(d)}
\put(-569,61){$C(\vartheta)$}
\put(-571,43){[$\mu\hbox{K}^2$]}
\put(-319,-78){$\vartheta$}
\put(-310,61){$C(\vartheta)$}
\put(-312,43){[$\mu\hbox{K}^2$]}
\put(-61,-78){$\vartheta$}
\put(-380,-40){$\alpha=1.4$}
\put(-380,-55){$\beta=1.0$}
\put(-121,-40){$\alpha=2.0$}
\put(-121,-55){$\beta=1.0$}
\vspace*{-25pt}
}
\end{center}
\caption{
\label{Fig:Correlation_General_Half_Turn_Space_Alpha}
The ensemble average of the temperature correlation function $C(\vartheta)$
for the general half-turn space for $\alpha = 0.5$, 0.7, 1.4 and 2.0
is shown.
The parameter $\beta$ is fixed as $\beta=1$.
The average over all positions of the observer is plotted as
a solid curve and its standard deviation as a dark grey band.
The distribution of the correlation function $C(\vartheta)$
depending on the position of the observer is given as a light grey band.
The dashed curve belongs to the position with the smallest value of
$S(60^\circ)$ and the dotted one to the largest value of $S(60^\circ)$.
}
\vspace*{-10pt}
\end{figure}

For the four cases presented in
figure \ref{Fig:S60_General_Half_Turn_Space_Observer},
the figure \ref{Fig:Correlation_General_Half_Turn_Space_Alpha}
displays the ensemble average of the correlation function $C(\vartheta)$.
The solid curve is the average over all positions of the observer,
and the dashed and dotted curves belong to the positions
at which the smallest and largest values of $S(60^\circ)$ occur.
Again one observes the trend of increasing variability of $C(\vartheta)$
with increasing values of $\alpha$.
This trend is also revealed by the increasing width of
the standard deviation, which is shown as a dark grey band.
The full width of variation is given by the light grey band,
which gives the maximal and minimal values of
the correlation function $C(\vartheta)$
that occur among the different positions
and shows the same trend.


\section{Comparison with Observations}

The discussion of the last section puts the focus on
the $S(60^\circ)$ statistic.
This quantity has the advantage that it is independent
of any measurements and describes the properties of the considered model.
In this section we compare the CMB properties of the half-turn space
with the correlation function $C^{\hbox{\scriptsize obs}}(\vartheta)$
obtained from the WMAP 7 year data.
We compute two correlation functions $C^{\hbox{\scriptsize obs}}(\vartheta)$.
The first one is obtained from the ILC 7 year map,
whereas the second one uses the same map
restricted to the pixels outside the KQ75 7yr mask
\cite{Gold_et_al_2010}.
Due to the recent discussions 
\cite{Copi_Huterer_Schwarz_Starkman_2008,Copi_Huterer_Schwarz_Starkman_2010,%
Efstathiou_Ma_Hanson_2009,Aurich_Lustig_2010}
on the relevance of these two correlation functions
we use both in the following analysis.
In order to compare the correlation function
$C^{\hbox{\scriptsize model}}(\vartheta)$
with the observed correlation function
$C^{\hbox{\scriptsize obs}}(\vartheta)$
the integrated weighted temperature correlation difference
\cite{Aurich_Janzer_Lustig_Steiner_2007}
\begin{equation}
\label{Eq:I_measure}
I := \int_{-1}^1 d\cos\vartheta \; \;
\frac{(C^{\hbox{\scriptsize model}}(\vartheta)-
C^{\hbox{\scriptsize obs}}(\vartheta))^2}
{\hbox{Var}(C^{\hbox{\scriptsize model}}(\vartheta))}
\end{equation}
is introduced which tests all angular scales $\vartheta\in[0^\circ,180^\circ]$.
The variance is computed using
\begin{equation}
\label{Eq:Var_C_theta}
\hbox{Var}(C(\vartheta)) \; \approx \;
\sum_l \frac{2l+1}{8\pi^2} \,  \left[C_l \,P_l(\cos\vartheta)\right]^2
\hspace{10pt} .
\end{equation}

\begin{figure}
\begin{center}
\vspace*{-30pt}
\begin{minipage}{11cm}
\includegraphics[width=10.0cm]{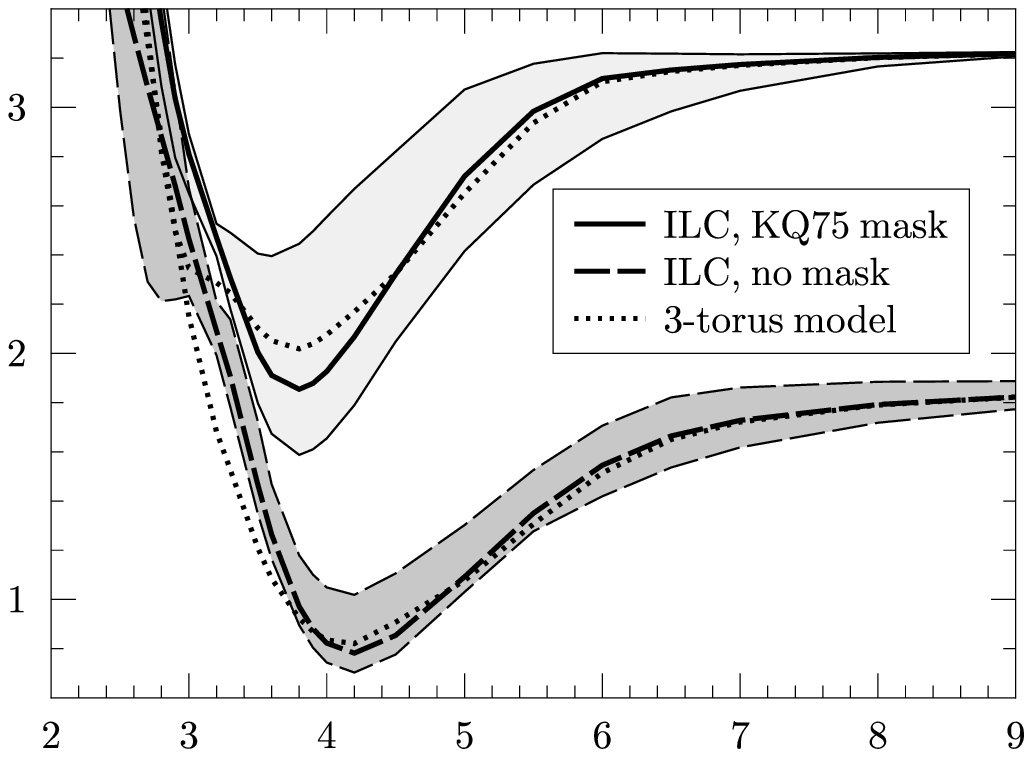}
\end{minipage}
\put(-60,-79){$L$}
\put(-305,55){$I(L)$}
\end{center}
\vspace*{-20pt}
\caption{\label{Fig:I_Cubic_Half_Turn_Space}
The integrated weighted temperature correlation difference $I(L)$
is shown for the cubic half-turn space depending on the
length $L$.
The results are shown for $C^{\hbox{\scriptsize obs}}(\vartheta)$
obtained from the ILC 7yr map with and without applying
the KQ75 7yr map.
The bands show the range of variation with respect to the
observer positions.
The dotted curves represent the corresponding results for the
cubic torus model which is a homogeneous space form
having no such range of variation.
}
\end{figure}

\begin{figure}
\begin{center}
\vspace*{-30pt}
\begin{minipage}{11cm}
\includegraphics[width=10.0cm]{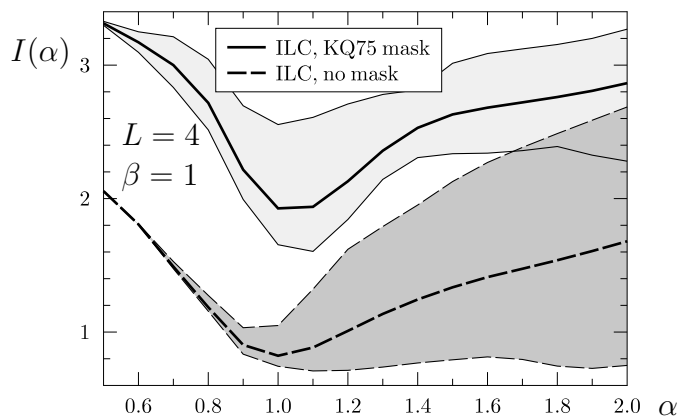}
\end{minipage}
\put(-60,-79){$\alpha$}
\put(-305,55){$I(\alpha)$}
\put(-263,23){$L=4$}
\put(-263,8){$\beta=1$}
\end{center}
\vspace*{-20pt}
\caption{\label{Fig:I_General_Half_Turn_Space}
The same quantities as in figure \ref{Fig:I_Cubic_Half_Turn_Space} are shown
but now for the general half-turn space with $L=4$ and $\beta=1$.
Thus, $I(\alpha)$ is plotted as a function of the distortion parameter $\alpha$.
}
\end{figure}

\begin{figure}
\begin{center}
\vspace*{-30pt}
\begin{minipage}{11cm}
\includegraphics[width=10.0cm]{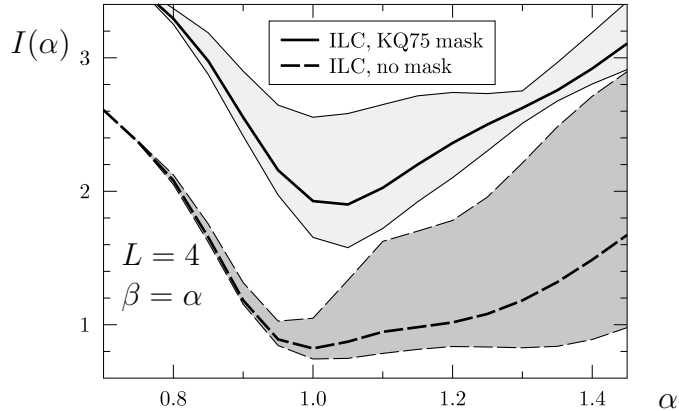}
\end{minipage}
\put(-60,-79){$\alpha$}
\put(-305,55){$I(\alpha)$}
\put(-263,-25){$L=4$}
\put(-263,-40){$\beta=\alpha$}
\end{center}
\vspace*{-20pt}
\caption{\label{Fig:I_General_Half_Turn_Space_diagonal}
The same quantities as in figure \ref{Fig:I_Cubic_Half_Turn_Space} are shown
but now for the general half-turn space with $L=4$ and $\beta=\alpha$.
Thus, $I(\alpha)$ is plotted as a function of the distortion parameter $\alpha$.
}
\end{figure}

The results are shown in figures
\ref{Fig:I_Cubic_Half_Turn_Space},
\ref{Fig:I_General_Half_Turn_Space}, and
\ref{Fig:I_General_Half_Turn_Space_diagonal}
for the three half-turn space sequences
that are studied in the previous sections.
The cubic half-turn space is parameterised by $L$,
and figure \ref{Fig:I_Cubic_Half_Turn_Space} reveals
that models with $L$ close to $L=4$ describe the data
better than the infinite volume concordance model
whose behaviour is approximately seen at $L=9>D_{\hbox{\scriptsize sls}}$.
It is worthwhile to note that the minimum close to $L=4$
is present in the comparison with the full ILC data set
as well as with the reduced data set by using the KQ75 7yr mask.
With the mask the minimum lies slightly below $L=4$ and
without slightly above.
This again justifies the restriction to models with
a volume ${\cal V} = L^3=64$ as discussed in the previous section.

The minimum around $L=2$ in the $S(60^\circ)$ statistic,
see figure \ref{Fig:S60_Cubic_Half_Turn_Space},
is now a local minimum with a value even above the $L=9$ value.
This is the reason why we do not discuss this volume ${\cal V}$
in detail in this paper.
But we would like to note that a further extension \cite{Liu_Li_2008a}
of the KQ75 7yr mask leads to a minimum at $L\simeq 2$ comparable to
that at $L\simeq 4$ as shown in \cite{Aurich_Lustig_Steiner_2009}.
Only future CMB data can decide whether this second minimum is
a genuine alternative to that at $L\simeq 4$.

The figure \ref{Fig:I_Cubic_Half_Turn_Space} also shows the
results for the homogeneous cubic torus model for the two
correlation functions $C^{\hbox{\scriptsize obs}}(\vartheta)$.
Although the values of $I(L)$ of the 3-torus are contained
within the range of variation of the half-turn space,
it is striking to see
that the minimum of the average over the observer positions
of the half-turn space is lower than that of the torus model.
Furthermore, the half-turn space provides observer positions
which possess an even better match with the observations
as revealed by the even lower values of $I(L)$.
Thus, the half-turn space describes the CMB data not only better than
the concordance model,
but even slightly better than the torus topology.

In figures \ref{Fig:I_General_Half_Turn_Space} and
\ref{Fig:I_General_Half_Turn_Space_diagonal}
the integrated weighted temperature correlation difference $I(\alpha)$
is shown as a function of the parameter $\alpha$
where the volume is fixed as ${\cal V} = 64$.
As in the case of the $S(60^\circ)$ statistic,
the range of variation with respect to the observer position increases
up to $\alpha=1$,
thereafter the behaviour is more involved and
increases only for the full ILC data set.
A preference for an almost cubic half-turn space occurs
when $I(\alpha)$ is computed by restricting the ILC data by the KQ75 mask.
Using the full ILC map leads to such an increase of the range of variation
that there are observer positions for $\alpha>1$ as good as
in the cubic case.
However, the average over the observer positions points to
a preference for a cubic half-turn space.
The preference for symmetrical space forms is also
found in other topologies \cite{Weeks_Luminet_Riazuelo_Lehoucq_2005}.


\section{Summary}

In this paper we study a model of our Universe
where the spatial space has a finite volume.
Although the cosmological parameters are those of the concordance model,
the finite spatial size leads to a suppression of the anisotropy 
in the CMB on large scales which is indeed observed in the data.
Most topological models that are studied in the literature,
have the property that they are homogeneous with respect to the
statistical properties of the CMB,
i.\,e.\ the statistical expectations are the same
for each observer position within the fundamental cell
when the ensemble average over the sky realisations is carried out.
This is different in the case of inhomogeneous space forms
where the comparison with the observational data is much more involved
since the variation of the observer position has to be taken into account.

The model system considered in this paper
is the so-called half-turn space form.
This inhomogeneous space tessellates the Euclidean space in a similar
way as the homogeneous 3-torus topology,
except that the identification of one pair of faces includes an
additional rotation with an angle of $180^\circ$.
This rotation leads to the inhomogeneity.
The large-scale angular power is conveniently described by the
temperature 2-point correlation function $C(\vartheta)$,
Eq.\,(\ref{Eq:C_theta}),
from which the $S(60^\circ)$ statistic, the scalar quantity
defined in Eq.\,(\ref{Eq:S_statistic_60}), can be obtained.
The latter is a measure of the power in the CMB anisotropies at
scales above $\vartheta \geq 60^\circ$
and has the advantage that it facilitates the comparison
of the various observer positions.
The figure \ref{Fig:S60_Cubic_Half_Turn_Space} shows
that cubic half-turn spaces with a topological length close to $L=4$
present a good choice with respect to the desired small power
on large angular scales.
The second minimum at $L=2$ is probably not favoured by the
current observations (see, however, \cite{Aurich_Lustig_Steiner_2009})
such that the following discussion puts the focus on spaces with $L=4$.
This leads to a volume ratio
${\cal V}_{\hbox{\scriptsize phys}}/{\cal V}_{\hbox{\scriptsize sls}}$
around 0.4 which is also favoured by several topological spaces
with positive curvature.
The figures \ref{Fig:S60_General_Half_Turn_Space} and
\ref{Fig:S60_General_Half_Turn_Space_diagonal} address the question
whether non-cubic half-turn spaces provide models with even lower
large-scale power.
Here, two sequences of asymmetric half-turn spaces are shown
which are parameterised by $\alpha$.
As can be seen in both figures, the average over all observer positions
gets its minimum close to the cubic half-turn space with $\alpha=1$.
However, in both cases there are observer positions
which possess also for $\alpha\gtrsim 1$ a large-scale power
almost as low as in the cubic case.
The dependence of the $S(60^\circ)$ statistic on the observer position
is visualised in figures \ref{Fig:S60_Cubic_Half_Turn_Space_Observer}
and \ref{Fig:S60_General_Half_Turn_Space_Observer}
which reveal the regions within the fundamental cell
that possess the desired small anisotropy.
This emphasises the variety of inhomogeneous space forms.

A comparison with the WMAP 7 year data \cite{Gold_et_al_2010}
is carried out using
the integrated weighted temperature correlation difference $I$
defined in Eq.\,(\ref{Eq:I_measure}).
The correlation function is computed from the full ILC 7yr map
as well as from this map again but subjected to the KQ75 7yr mask.
For both correlation functions the figure \ref{Fig:I_Cubic_Half_Turn_Space}
shows the result for the cubic half-turn space.
The minimum in $I(L)$ around $L=4$ is clearly revealed,
and it demonstrates that not only the low power on large angular scales
as expressed by the $S(60^\circ)$ statistic favours this size
for the fundamental cell, but also the direct comparison of the
corresponding correlation functions as in Eq.\,(\ref{Eq:I_measure}).
The figure \ref{Fig:I_Cubic_Half_Turn_Space} presents also the result
for the cubic torus model which is a homogeneous space form.
It shows that most observer positions of the half-turn space
provide a slightly better description of the CMB data
than the 3-torus topology.
The figures \ref{Fig:I_General_Half_Turn_Space} and
\ref{Fig:I_General_Half_Turn_Space_diagonal} display the
results for the two sequences of asymmetric half-turn spaces
parameterised by $\alpha$.
These figures reveal that the special case of the cubic half-turn space
yields the best description,
although there are observer positions for $\alpha\gtrsim 1$
that describe the observed correlation function almost equally well.
This is consistent with the result obtained from the $S(60^\circ)$ statistic.

The analysis of this paper shows that the simplest inhomogeneous
flat topology describes the large-scale angular anisotropy of the CMB
better than the $\Lambda$CDM concordance model.
Although the agreement is even slightly better than those of
the flat 3-torus model,
the concrete identification of the topology requires a more direct
measure of the topological signal as given by,
e.\,g.\ the spatial correlation function
\cite{Roukema_et_al_2008a,Aurich_2008} or
the covariance matrix
\cite{Phillips_Kogut_2006,Kunz_et_al_2006,Kunz_et_al_2007,%
Aurich_Janzer_Lustig_Steiner_2007}.
But this is left to a future work.


\section*{Acknowledgements}

We would like to thank the Deutsche Forschungsgemeinschaft
for financial support (AU 169/1-1).
HEALPix (healpix.jpl.nasa.gov)
\cite{Gorski_Hivon_Banday_Wandelt_Hansen_Reinecke_Bartelmann_2005}
and the WMAP data from the LAMBDA website (lambda.gsfc.nasa.gov)
were used in this work.
The computations are carried out on the Baden-W\"urttemberg grid (bwGRiD).


\section*{References}

\bibliography{../bib_astro}

\bibliographystyle{h-physrev5}

\end{document}